\documentclass[journal]{IEEEtran}
\usepackage{amsmath,amsfonts}
\usepackage{algorithmic}
\usepackage{algorithm}
\usepackage{array}
\usepackage[caption=false,font=normalsize,labelfont=sf,textfont=sf]{subfig}
\usepackage{textcomp}
\usepackage{stfloats}
\usepackage{url}
\usepackage{verbatim}
\usepackage{graphicx}
\usepackage{cite}
\usepackage{color}
\usepackage{hyperref}
\usepackage{booktabs} 
\usepackage{tabularray}
\usepackage{multirow}
\usepackage[normalem]{ulem}
\usepackage{xcolor}
\usepackage{ulem}
\usepackage{makecell}
\usepackage{graphicx} 
\usepackage{xcolor}
\usepackage{tikz}

\newcommand{\toolName}[1]{\textit{COIVis}}

\newcommand{\edit}[1]{#1}
\newcommand{\delete}[1]{}

\newcommand{\viewname}[1]{\textbf{#1}}

\newcommand{\feature}[1]{\textit{#1}}

\begin{document}
\title{\toolName{}: Eye-tracking-based Visual Exploration of Concept Learning in MOOC Videos}

\author{
Zhiguang Zhou, Ruiqi Yu, Yuming Ma, Hao Ni, Guojun Li, Li Ye, Xiaoying Wang,\\
Yize Li, Yigang Wang, and Yong Wang
\thanks{Zhiguang Zhou, Ruiqi Yu, Guojun Li, Yize Li, and Yigang Wang are with Hangzhou Dianzi University, Hangzhou, China. E-mail: \{zhgzhou, richyu, liyize, yigang.wang\}@hdu.edu.cn; lgj\_afeng@163.com.}%
\thanks{Yuming Ma is with Xi'an Jiaotong University, Xi'an, China. E-mail: mamingming@stu.xjtu.edu.cn.}%
\thanks{Hao Ni and Li Ye are with Zhejiang University, Hangzhou, China. E-mail: 529334189@qq.com; li-ye@zju.edu.cn.}%
\thanks{Xiaoying Wang is with the University of Southern California, Los Angeles, CA, USA. E-mail: delluluwxy@gmail.com.}%
\thanks{Yong Wang is with Nanyang Technological University, Singapore. E-mail: yong-wang@ntu.edu.sg.}%
\thanks{(Corresponding authors: Yize Li and Yong Wang.)}%
}

\markboth{IEEE TRANSACTIONS ON VISUALIZATION AND COMPUTER GRAPHICS, VOL. xx, NO. x, JUNE 20xx}%
{Shell \MakeLowercase{\textit{et al.}}: A Sample Article Using IEEEtran.cls for IEEE Journals}

\maketitle


\begin{abstract}
Massive Open Online Courses (MOOCs) make high-quality instruction accessible. However, the lack of face-to-face interaction makes it difficult for instructors to obtain feedback on learners' performance and provide more effective instructional guidance. Traditional analytical approaches, such as clickstream logs or quiz scores, capture only coarse-grained learning outcomes and offer limited insight into learners' moment-to-moment cognitive states. In this study, we propose \toolName{}, an eye-tracking-based visual analytics system that supports concept-level exploration of learning processes in MOOC videos. \toolName{} first extracts course concepts from multimodal video content and aligns them with the temporal structure and screen space of the lecture, defining Concepts of Interest (COIs), which anchor abstract concepts to specific spatiotemporal regions. Learners' gaze trajectories are transformed into COI sequences, and five interpretable learner-state features---Attention, Cognitive Load, Interest, Preference, and Synchronicity---are computed at the COI level based on eye tracking metrics. Building on these representations, \toolName{} provides a narrative, multi-view visualization enabling instructors to move from cohort-level overviews to individual learning paths, quickly locate problematic  concepts, and compare diverse learning strategies. We evaluate \toolName{} through two case studies and in-depth user-feedback interviews. The results demonstrate that \toolName{} effectively provides instructors with valuable insights into the consistency and anomalies of learners' learning patterns, thereby supporting timely and personalized interventions for learners and optimizing instructional design. 
\end{abstract}






\graphicspath{{figs/}{figures/}{pictures/}{images/}{./}} 




\section{Introduction}\label{sec:intro}

\begin{figure*}[ht]
\centering
\includegraphics[width=\linewidth]{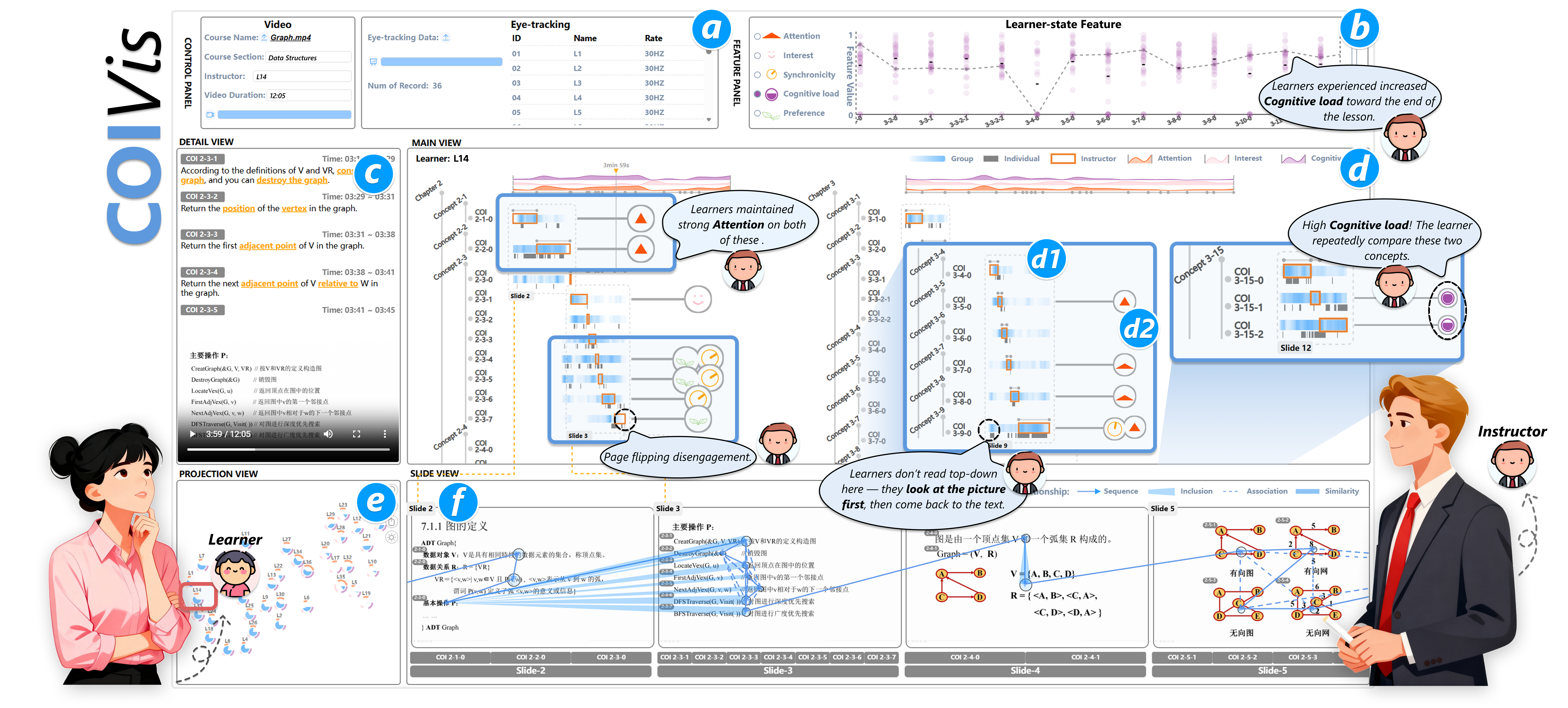}
\caption{An instructor uses \toolName{} to explore learners' performance in the MOOC video \textit{Definitions and Terminology of Graphs} from a COI (Concept of Interest) perspective. The video and eye tracking data are imported via the Control Panel (a), processed in the backend, and displayed in the Detailed View (c), which shows the instructor’s explanation of each COI along with the original video. The Main View (d) shows all learners’ interactions with the COIs (d1), while the Feature Panel (b) filters and highlights key COIs using bubbles and icons (d2). Individual learner data can be selected in the Projection View (e). The Slide View (f) breaks down the video content by COI, aligning with the Main View to provide contextual semantic supplementation.}
\label{main}
\end{figure*}

\IEEEPARstart{I}{n} recent years, the proliferation of online learning platforms such as Massive Open Online Courses (MOOCs) has made high-quality educational resources accessible worldwide, breaking traditional temporal and spatial constraints. However, these advances have been accompanied by poor learning persistence and high dropout rates~\cite{onahDropoutRatesMassive2014,betaitiaExploringDropoutRates2025,cismasuPersistencePuzzleBibliometric2025}. Because MOOCs typically rely on one-way video lectures, it is difficult for instructors and learners to communicate and adjust the teaching process in a timely manner, as in traditional face-to-face classrooms~\cite{nujidReviewEngagementStrategies2024,yousefEffectiveDesignBlended2015,zhangAnalysisCoresThese2020}. Existing online learning platforms primarily employ quizzes~\cite{koedinger2015learning} and educational data mining techniques such as clickstream analysis~\cite{wei2020predicting} for assessment, which often lack real-time responsiveness and can be superficial, subjective, and limited in reliability and validity~\cite{raiGeneralImpactMOOC2017, taoConstructionResearchEvaluation2024}. This calls for assessment approaches that can continuously and accurately reflect learners’ states during the learning process.

We address this gap by defining learners’ gaze areas as COIs (Concepts of Interest), grounded in a semantic understanding of concepts, to provide instructors with process-oriented, concept-level evidence of learners’ understanding. Building on this notion, we develop a visual teaching aid that integrates eye-tracking data with COIs in MOOC videos to offer more objective feedback. In a smart classroom where each workstation is equipped with an eye tracker, we collect real-time gaze data from all learners during MOOC viewing sessions. Based on in-depth interviews with experienced instructors, we identify three key challenges for MOOC learning assessment. First, concepts are implicitly embedded in rich multimodal information; identifying learners’ COIs requires structural parsing of video frames and organizing COI representations as a basis for assessing conceptual performance \textbf{(CH1)}. Second, instructors must understand which aspects of learners’ behavior can meaningfully inform course improvement, yet intricate eye-tracking metrics must be translated into insights that align with instructors’ expectations \textbf{(CH2)}. Third, traditional MOOC videos are designed from usability or learning-theory perspectives and lack mechanisms tailored to learners and their cognitive patterns, so systems should enable instructors to observe class-wide learning, rapidly locate specific learners or content segments, and optimize instructional design \textbf{(CH3)}.

To address these challenges, we develop \toolName{}, an exploratory visual analytics system that uses eye tracking as a proxy for learners’ cognitive processing and introduces COIs tailored to MOOC scenarios. Concepts are extracted from MOOC videos using video processing and speech recognition, and their alignment with the screen space defines COIs; learners’ gaze trajectories are represented as sequences over COIs, characterizing their learning processes. A workshop with eight experienced instructors clarifies which aspects of learners’ engagement they prioritize. Drawing on these requirements and an extensive literature review, we compute five per-COI learner-state features---\feature{Attention}, \feature{Cognitive Load}, \feature{Interest}, \feature{Preference}, and \feature{Synchronicity}. We encode these features in a narrative visualization that presents learners’ learning processes in MOOC videos. \toolName{} offers rich interactions, providing both group- and individual-level overviews while enabling instructors to examine specific learners or COIs in detail; two case studies and a user-feedback interview demonstrate the effectiveness and usefulness of our system.

The contributions of this paper are summarized as follows:

\begin{itemize}
  \item We employ eye tracking to transform learners’ learning processes in MOOC videos into COI (Concept of Interest) sequences enriched with temporal, sequential, and semantic information.
  \item We identify representative learner-state features of learners during the learning process through literature review and user interviews, and quantify them using eye tracking metrics at the COI level.
  \item We propose an interactive system, \toolName{}, to assist instructors in exploring learners’ processes of absorbing and processing concepts, uncovering intentional patterns to foster personalized teaching and optimize instructional designs.
\end{itemize}

\section{Related Work}\label{sec:rw}
We review research related to our work: visual analytics of online learning data, eye-tracking-based visual analytics in education, and multimodal, semantics-driven visual analytics.
    
\subsection{Visual Analytics of MOOC Data}\label{sec:moocva}
The popularity of MOOCs and other online learning platforms has generated large-scale behavioral logs, creating rich opportunities for visual analytics of learning processes. Many systems rely on clickstream data to characterize viewing and interaction patterns. For example, VisMOOC visualizes seek operations and other video interactions through coordinated views to help instructors understand learners’ viewing habits~\cite{shi2015vismooc}, PeakVizor uses glyphs, flow maps, and correlation views to reveal “peaks” in video interaction and explore group differences and anomalies~\cite{chen2015peakvizor}, and sequence-based approaches such as ViSeq compare learning paths across learner groups~\cite{chen2018viseq}. These systems offer useful insights into global trends and key interaction points and have informed the design of targeted interventions.

However, many approaches focus on event counts or statistical patterns, making it difficult to capture deeper cognitive processes and self-regulation, and clickstream data are inherently noisy and fragmentary for robust cognitive inference~\cite{baker2020benefits}. Visualizations are often organized around video control events or coarse interface regions with limited alignment to semantic knowledge units, which hinders direct interpretation of learners’ conceptual processing. Furthermore, integrating multimodal content structures (e.g., video semantics, textual materials, assessments) with learner behavior in a unified visual analytics framework remains challenging~\cite{jeon2020dropout}.

\subsection{Eye Tracking in Online Learning}\label{sec:eyetracking} 
Eye tracking has been widely used in educational research \edit{for understanding how learners allocate visual attention and process information in online learning environments.}
Surveys show that heatmaps, scanpaths, fixation density plots\edit{, and attention maps} remain dominant visualization forms for summarizing attention patterns~\cite{alemdag2018systematic, el2019eye}\edit{,~\cite{blascheck2017visualization}}, and dashboard-style systems such as VAAD use these data to compare group- and individual-level behaviors and support predictive modeling~\cite{navarroVAADVisualAttention2024}.

\edit{Taken together, existing surveys of eye tracking in online and multimedia learning characterize the field as methodologically mature on the visualization side but analytically narrow on the interpretive side~\cite{alemdag2018systematic, el2019eye, blascheck2017visualization}. The prevailing outputs are screen-region-centric attention summaries, and the diagnostic potential of gaze data for instructors, particularly at the level of the content concepts they actually teach, has received far less attention.}

\edit{Despite these advances, two limitations persist. First, most eye tracking visualizations remain anchored to screen-region AOIs and do not align gaze-based attention with the instructional content structure, making it difficult for instructors to derive concept-level diagnostic insights. Second, while individual metric--state mappings are well documented, they are typically studied in isolation at the trial or page level; no existing framework integrates them into a unified, multi-dimensional characterization of learner engagement at the concept level.}

\subsection{Multimodal Data Analysis}\label{sec:multimodal}
Multimodal and semantics-driven visual analytics has become an important direction in educational visualization~\cite{wang2024multi}. Some work focuses on multimodal analysis of learning behaviors: for example, Immadisetty et al.~\cite{immadisettyMultimodalityOnlineEducation2023} combine posture, gestures, facial expressions, eye tracking, and speech to recognize learners’ emotions and engagement, illustrating the benefits of fusing multiple channels. Other research centers on semantic concept modeling and visualization of knowledge structures, such as ConceptThread, which automatically extracts core concepts and their hierarchical and temporal relations from MOOC videos and visualizes them with a thread metaphor to support understanding of course structure~\cite{zhou2024conceptthread}. A third line constructs multimodal knowledge graphs and time-sensitive representations; for instance, TIVA-KG unifies text, images, video, and audio into a knowledge graph with triplet grounding~\cite{wang2023progressive}, and TimeChat uses a time-aware frame encoder and a sliding video Q-Former for improved temporal localization and semantic alignment in long videos~\cite{renTimeChatTimesensitiveMultimodal2024}. Related techniques such as temporal video pyramids~\cite{swift2022visualizing} and lecture video segmentation via action recognition and speech analysis~\cite{imran2012exploiting} further enrich the technical basis for cross-temporal, cross-modal analytics.

Even so, concept-level analysis of learning processes remains underexplored. Many multimodal systems focus on video content or global trends (e.g., emotion recognition, frame–audio fusion) without detailing how learners process individual concepts, and concept visualization is rarely aligned with learners' multimodal behavior data. Moreover, many systems provide static or aggregated results, limiting support for interactive exploration and instructor-driven diagnosis. 

\section{User-Centered Design}\label{sec:ucd}
\edit{To identify learner states for concept-level learning analysis and determine which of them matter most to instructors,} 
we followed a user-centered design paradigm, combining a literature review with semi-structured interviews of frontline instructors to derive concrete system requirements. 
\edit{Specifically, Section~\ref{sec:litreview} surveys the literature to compile a pool of learner-state constructs amenable to eye-tracking measurement, and Section~\ref{sec:interview} then grounds these constructs through instructor interviews, yielding the design considerations and requirements that guide the system.}

\subsection{Literature Review}\label{sec:litreview}
\edit{To identify which learner states can be captured through eye tracking at the concept level,}
we conducted a systematic review of work on online learning, behavioral analysis, and eye tracking visualization from 2010–2025 across major education and visualization venues, complemented by Google Scholar searches. 
\edit{Beyond venue- and keyword-based retrieval, we traced forward and backward citations across core authors' lines of work, not isolated papers.}
We focused on studies involving MOOC or online video learning, process data such as eye tracking, and concept- or knowledge-level content organization.In total, 67 peer-reviewed papers were retained and coded along four dimensions—research domain, data modality, analysis method, and analysis task—as summarized in Table~\ref{tab:litreview}.

\begin{table}[tb]
  \centering
    \caption{Distribution of coded categories for the 67 reviewed papers. Each paper may belong to multiple categories.}
  \label{tab:litreview}
  \setlength{\tabcolsep}{4pt}       
  \renewcommand{\arraystretch}{1.05}
  \footnotesize                    
    \begin{tabular}{p{0.35\linewidth}@{\hspace{2pt}}c
                    p{0.35\linewidth}@{\hspace{2pt}}c}

    \toprule
    \multicolumn{2}{c}{\textbf{Research Domain}} 
      & \multicolumn{2}{c}{\textbf{Data Modality}} \\
    Category & Count & Category & Count \\
    \midrule
    Education      & 48 & Concept     & 43 \\
    Visualization  & 31 & Performance & 37 \\
    Algorithm      & 23 & Area        & 26 \\
    Interaction    & 20 & Text        & 24 \\
    Psychology     & 15 & Video       & 20 \\
    Others         & 15 & Image       & 16 \\
    Statistics     & 14 & Others      & 14 \\
                   &    & Audio       & 11 \\
    \midrule
    \multicolumn{2}{c}{\textbf{Analysis Method}} 
      & \multicolumn{2}{c}{\textbf{Analysis Task}} \\
    Category & Count & Category & Count \\
    \midrule
    Test           & 41 & Attention           & 32 \\
    Eye tracking   & 33 & Cognitive load      & 27 \\
    Education data & 27 & Interest            & 16 \\
    Clickstream    & 18 & Preference          & 15 \\
    Thinking aloud & 12 & Others              & 14 \\
    Face detection &  8 & Cognitive style     & 11 \\
    Others         &  8 & Synchronicity       &  8 \\
    EEG            &  6 & Satisfaction        &  7 \\
                   &    & Cognitive crossover &  5 \\
    \bottomrule
  \end{tabular}
\end{table}

\textbf{Concept-based Analysis.}
The literature indicates that a core issue in online learning and educational data mining research is selecting the appropriate unit of analysis. Some works aggregate learning performance at the course or chapter level to assess overall learning outcomes~\cite{papamitsiou2014learning,viberg2018current,zhu2022trends}, while others analyze behavioral patterns based on video segments or activity sequences~\cite{shi2015vismooc,chen2015peakvizor,sinha2014your}. In recent years, an increasing number of studies have attempted to refine the granularity to the level of knowledge points or concepts~\cite{liu2018conceptscape,barria2017concept,zhu2022trends}, particularly in video-based online courses~\cite{balasubramanian2016multimodal,ghosh2022augmenting,zhou2024conceptthread}. In such contexts, course content is typically organized as a series of interconnected concepts, and the learning process is reflected in the learner's progressive encoding, revisiting, and integration of these concepts. Relevant research suggests that analyzing learning at the concept level helps reveal which knowledge points present difficulties and how learners transfer and integrate information across concepts.

\textbf{Eye-tracking-based Analysis.}
Regarding data modalities, considerable online learning research relies on quiz scores, clickstream logs, and forum posts to analyze learning behavior~\cite{sinha2014your,papamitsiou2014learning,littenberg2020studying}. While suitable for assessing outcomes or macro-level patterns, these data sources are often indirect for characterizing finer-grained temporal and spatial learning processes. In contrast, eye tracking can continuously record learners' gaze positions and temporal trajectories on the screen without interrupting learning tasks. Consequently, it is increasingly used in contexts such as multimedia learning, programming, and visual analytics to study attention allocation and cognitive processes~\cite{van2010eye,alemdag2018systematic,chytry2025using}. A typical approach involves dividing stimuli into Areas of Interest (AOIs) and analyzing metrics such as fixation duration, fixation count, and regression count within each AOI. Concurrently, research has proposed various eye tracking visualization and multimodal analysis methods to display gaze paths and dynamic attention distributions~\cite{andrienko2012visual,burch2018eye}. Collectively, these works demonstrate that eye tracking data can serve as critical process data bridging ``how content is presented'' and ``how learners visually engage.''

\textbf{Learner-state Constructs.}  
Beyond units of analysis and data modalities, researchers focus on which psychological and behavioral indicators reflect the learning process. In educational psychology and learning analytics, researchers have proposed numerous learner-state constructs to describe learning states, such as attention~\cite{posner2007research,di2010saccadic,bachurina2022multiple}, cognitive load~\cite{sweller2011cognitive,van2010cognitive}, interest~\cite{hidi2006four,renninger2015power}, motivation~\cite{pintrich1999role}, and engagement~\cite{fredricks2004school,henrie2015measuring}. Based on a systematic review and coding of relevant studies, we compiled over 20 learner-state constructs widely used in online and multimedia learning contexts, along with their common behavioral indicators. This resulted in a ``learner-state construct pool'' summarizing the primary dimensions of learning states focused on by existing research. 

According to our literature review, current research on online learning analytics remains at a macro, coarse-grained level, making it difficult to pinpoint concepts that interest learners or where they encounter comprehension barriers. By aligning eye tracking data with the conceptual structure of the course and providing evidence to instructors via visual analytics, this study aims to elevate online learning analysis to the concept level. This approach promises more targeted insights and a viable path for refined analysis.

\subsection{User Interview}\label{sec:interview}
In Section\edit{~\ref{sec:litreview}}, we compiled over 20 common learning behavior and psychological constructs from educational psychology and learning analytics, forming an initial ``construct pool.'' However, these constructs reflect states prioritized from a researcher's perspective and may not represent the online learning performance most relevant to instructors in real teaching contexts.

To address this, we conducted in-depth semi-structured interviews with instructors (P1–P8) to gain qualitative insights. P1–P5 primarily teach computer science, P6 teaches mathematics, and P7–P8 teach art and design-related courses. All participants have online teaching experience; specifically, P1, P2, and P7 have offered online course videos on platforms like Coursera and Chinese University MOOC. During the interviews, we guided participants to: (1) describe their process of developing teaching materials online or offline; (2) share challenges encountered in conducting online courses; and (3) summarize the aspects of learner performance they are most concerned with in online teaching and rank the importance of these concerns. 

\textbf{Learner-centered Materials.} When developing teaching materials, instructors often rely heavily on their own teaching experience and personal preferences. This was particularly evident in P1's statement: \textit{"Most of my slides and assignments come from problems I find interesting myself. I rarely have the chance to assess whether these are challenges learners actually enjoy solving."} P4 shared a similar sentiment: \textit{"I tend to use examples and case studies that inspire me personally, assuming they will resonate with learners in the same way."} Through the interviews, we found that instructors universally desire to understand which concepts and examples genuinely stimulate learners' intrinsic interest, such as stable learner preferences for concept media forms (images vs.\ text).


\textbf{Urgent Need for Feedback.} All interviewed instructors emphasized the difficulty of grasping learners' learning states in a timely and fine-grained manner in online contexts. P6 noted: \textit{"The intuitive understanding derived from learners' expressions and body language in face-to-face classrooms disappears online. I can't observe lapses in attention, making it hard to extend attention-based adaptive tutoring."} P5 also stated: \textit{"I often have to verbally ask if everyone is following, but usually only the top learners respond, while the majority remain silent."} Through systematic coding and salience scoring of the interview responses, we found that instructors generally feel that lacking the spontaneous and subtle feedback cues of face-to-face classrooms forces them to rely on coarse-grained indicators like grades or sparse comments. This indicates an urgent need among instructors for more timely and granular feedback on learning states.


\textbf{Representative learner-state features.}
In the interviews, instructors emphasized their desire to understand learner learning behaviors through various means. Some directly pointed out specific metrics, while others expressed concerns indirectly through classroom scenarios and teaching difficulties. Based on salience scoring of the coded interview data, we initially identified seven learner-state features: \feature{Attention}, \feature{Cognitive Load}, \feature{Interest}, \textit{Retention}, \textit{Cognitive Crossover}, \feature{Preference}, and \feature{Synchronicity}.

Among these, we further filtered out two features with lower feasibility for implementation. \textit{Retention} primarily involves knowledge maintenance over a longer time scale, which is difficult to reliably capture via eye tracking behavior within the short time window of a single concept. \textit{Cognitive Crossover} often involves relationships between multiple conceptual units and is more suitable as a high-level feature across Concept of Interest (COI) segments rather than being estimated independently on each COI. Therefore, in the current system implementation, we focus on five learner-state features that are both high-frequency and operable at the concept level: \feature{Attention}, \feature{Cognitive Load}, \feature{Interest}, \feature{Preference}, and \feature{Synchronicity}. Their operational definitions are as follows:

\begin{itemize}
  \item \feature{Attention}: Refers to the degree of sustained focus on task-relevant information. Instructors care first about who is focused and who is distracted, as this directly affects concept mastery and retention.
  \item \feature{Cognitive Load}: Refers to the occupation of working memory resources by learning tasks. Instructors want to identify which concepts cause learners to get ``stuck'' or ``fall behind'' to adjust difficulty and pacing.
  \item \feature{Interest}: Refers to the learner's sustained curiosity and intrinsic motivation toward specific content or presentation forms. Instructors care about which concepts and examples truly ``attract'' learners, rather than just eliciting superficial quietness.
  \item \feature{Preference}: Refers to stable preferences for content types and presentation modes (formulas, text, diagrams, cases, etc.). Instructors want to know which forms and concepts learners are more willing to invest time in.
  \item \feature{Synchronicity}: Refers to the consistency between the learning pace and the teaching schedule. Instructors pay attention to who is keeping up and who is falling behind, evidenced by behaviors like frequent pausing, re-watching, or significant lag.
\end{itemize}

We then group these features into two functional categories:
\begin{itemize}
  \item \textbf{Intra-concept engagement}: The first three features characterize the learner's internal response to a single concept.
  \item \textbf{Inter-concept integration}: The latter two emphasize the match between learner behavior and instructional design intent as well as pacing consistency.
\end{itemize}

These five features, rooted in instructors' actual concerns, provide a clear behavioral framework for the design of metrics based on eye tracking and behavioral data in Section\edit{~\ref{sec:coi}}.

\subsection{Design Requirements}\label{sec:designreq}

Focusing on the analysis of concepts (object) and learners (subject) has been widely recognized as crucial for enhancing the effectiveness of online learning. Guided by this core perspective and interviews with instructors, we collaboratively derived a set of specific design requirements \textbf{(R)} from three high-level design considerations \textbf{(DC)}. These requirements collectively aim to facilitate in-depth exploration of learners' concept-level learning processes.

\vspace{1pt}
\textbf{DC1: Connect learners with concepts in videos.}

\textbf{R1. Extract and align concepts from multimodal MOOC videos. }
Accurately aligning videos with their embedded concepts is crucial for constructing COIs. Due to the multimodal nature of video content—speech, text, and graphics—this requires advanced cross-modal semantic integration. Effective alignment enables meaningful COI mapping, providing a structured foundation for analysis and precise interpretation of conceptual representations in learning materials.

\textbf{R2: Map learners' eye tracking data to COIs.}  
P4 stated, \textit{"I am concerned with how to determine whether learners have viewed this concept."} To connect learners with course concepts, it is essential to map their eye tracking coordinates to the corresponding COIs. By treating the screen as an interactive medium, learners' gaze points align with COIs, converting raw eye-movement data into meaningful information on concept-level engagement. This alignment reveals which concepts learners focus on and the duration of their attention.

\vspace{1pt}
\textbf{DC2: Interpret learners' behaviors on COIs.}

\textbf{R3. Characterize learners' spatiotemporal learning path. }
Understanding learners’ engagement with video content requires analyzing their spatiotemporal learning path, which captures the dynamic interplay between where and when attention is directed. \textit{"We need to know not only where learners are focusing but also when they watch, for how long, and in what sequence,"} P2 asserted. The goal is to track learners in dealing with concepts, including detours, revisits, and even uncovering eye movement patterns.

\textbf{R4. Capture learners' high-level learner-state features.}
Tracking learners' learning paths reveals where they look but fails to capture the depth and quality of their conceptual engagement. High-level learner-state features offer deeper insights by reflecting learners' cognitive and behavioral states during concept interaction. These features go beyond basic navigation, incorporating physiological responses such as attention and cognitive load, as well as the alignment between learners' learning paths and instructional design.

\vspace{1pt}
\textbf{DC3: Support instructor-driven insight and guidance.}

\textbf{R5. Prioritize instructionally significant COIs for targeted insights.}
Instructors emphasized that learner performance relies on both individual abilities and concept features. The variability in outcomes underscores the need to identify which COIs warrant closer attention. As P1 noted, \textit{"Simply observing learners' performance on concepts is not enough; I need to understand what these features mean and how they impact learning outcomes."} P8 echoed this, stressing the importance of highlighting concepts where learners face cognitive challenges, disengagement, or performance shifts.

\textbf{R6. Target learners of interest for in-depth exploration. }
Instructors often focus on specific learners whose behaviors stand out due to unique learning paths, challenges, or achievements. While class data shows general trends, individual learners offer insights not always apparent at the group level. P4 and P5 advocated for a system examining individual performance on specific concepts, transitioning from aggregated data to detailed learner behaviors, ultimately enhancing teaching strategies for diverse learning experiences.

\section{COI}\label{sec:coi}
With the design requirements established in Section\edit{~\ref{sec:ucd}}, we introduce \emph{Concepts of Interest (COIs)} and construct a unified analysis framework that links MOOC video content with learners' eye tracking behavior for subsequent sequence modeling and learner-state feature computation.

\subsection{Concept and COI}\label{sec:concept}
Analyzing learning at the concept level requires clarifying how concepts function within a course. We treat a \emph{concept} as a self-contained knowledge unit, structured by the course’s chapters, headings, and discourse into themes and sub-concepts (e.g., ``DFS,'' ``loop'')~\cite{novak2008theory,zhou2024conceptthread}. Relations such as sequence, inclusion, association, and similarity link these concepts into a course-level concept map, which is then aligned with the temporal axis of instructional segments to characterize the knowledge structure and its presentation in the video.

In eye tracking analysis, \emph{Areas of Interest (AOIs)} aggregate gaze behavior by defining screen regions~\cite{alemdag2018systematic,el2019eye,holmqvist2011eye}. 
However, AOIs based only on spatial layout (e.g., ``title area'' or ``left image panel'') may not align with the concepts instructors care about. 
Instead, we use the concept map as a framework and annotate all visual elements associated with a concept—text, formulas, diagrams, code—as concept-driven AOIs within each instructional segment.

On this basis, we define a \emph{Concept of Interest (COI)} as a concept-indexed spatiotemporal region of the learning material that can serve as a unit for aggregating learners' eye tracking evidence. Each COI is specified by three components: (1) \textbf{Concept Identifier}, links each COI to a specific node in the concept map, thereby tracking which concept is being studied; (2) \textbf{Instructional Episode}, captures the spatiotemporally aligned video, audio, and slide intervals during which the instructor presents this concept along the lecture timeline; and (3) \textbf{Concept-driven AOIs}, define the COI's visual footprint as the set of on-screen visual elements that instantiate the concept during its instructional episode. Subsequently, learners' eye tracking events are mapped to COIs: a sample is assigned if its timestamp lies within the instructional episode and its screen position falls inside any concept-driven AOI. COIs thus provide concept-centered spatiotemporal regions for jointly organizing teaching episodes and learners' eye tracking evidence into semantically coherent units of analysis.

\begin{figure}[t]
  \centering
  \includegraphics[width=\columnwidth]{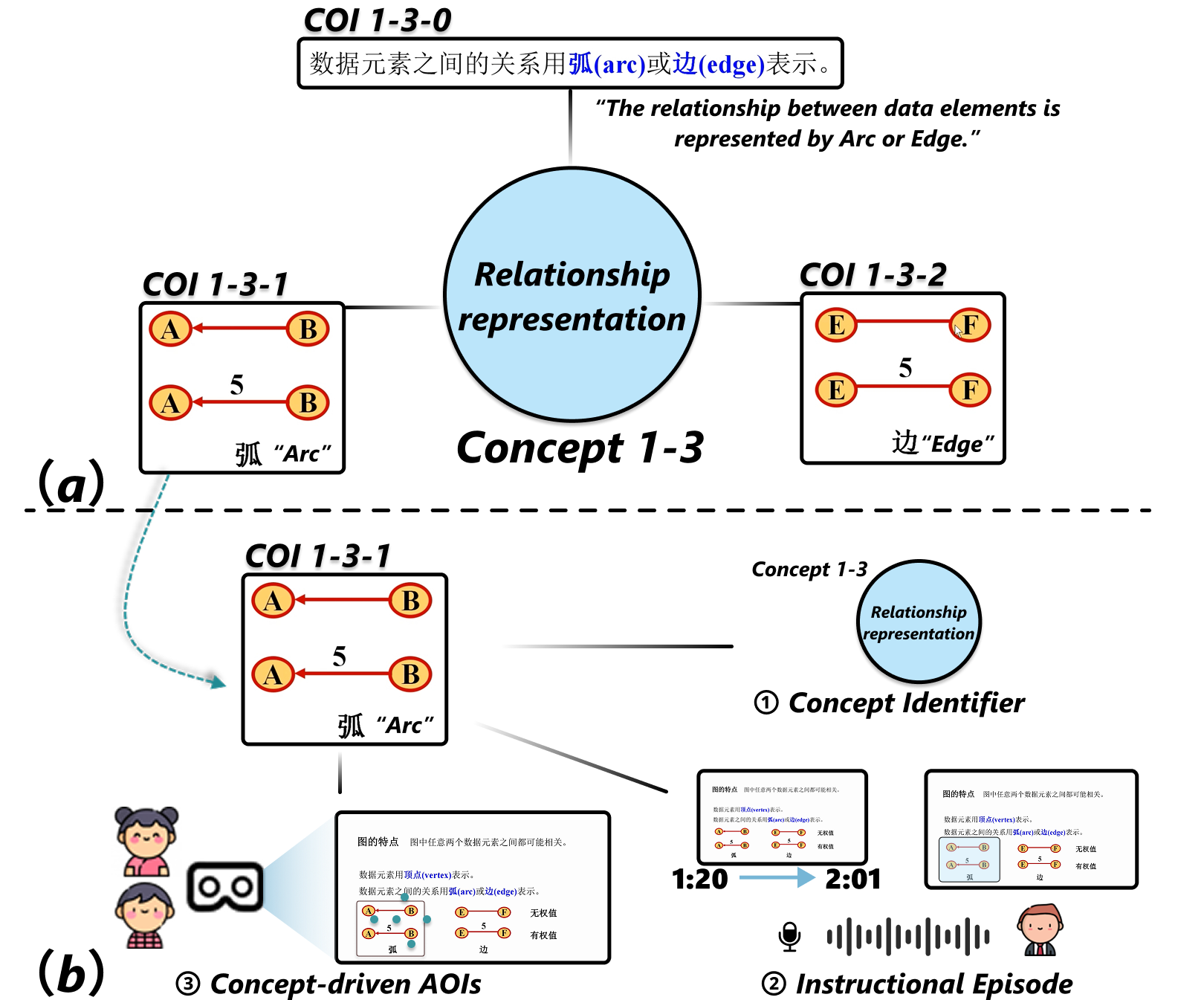}
\caption{Illustration of the COI notion using Concept~1--3 ``Relationship representation'': (a) the textual definition (COI~1--3--0) and two diagrams (COI~1--3--1/2) as concept-driven AOIs within one instructional episode; (b) COI~1--3--1 illustrating the three components of a COI: the concept identifier, instructional episode (slide interval and narration), and the concept-related gaze trace aggregated within the AOI.}
  \label{fig:concept-coi}
\end{figure}

To support analysis at multiple granularities, we use a hierarchical indexing scheme for COIs that follows the course structure. In the basic case, a COI is indexed as ``COI~$x$--$y$--$z$,'' where $x$ denotes the chapter index, $y$ the concept index within that chapter, and $z$ a specific COI instance associated with that concept. When finer-grained structural levels (e.g., sub episodes or nested concept-driven AOIs) need to be distinguished, we extend this notation with an additional index and write ``COI~$x$--$y$--$z$--$w$''; the first three indices still specify the chapter and concept context, and $w$ enumerates more fine-grained COIs within that context. For example, in Fig.~\ref{fig:concept-coi}, the concept ``Relationship representation'' is indexed as Concept~1--3; the textual definition and two illustrative examples correspond to COI~1--3--0, COI~1--3--1, and COI~1--3--2, respectively. These three COIs are independent instances at the implementation level but all belong to Concept~1--3.

Building on concept-level relations, we further define relations between COIs to characterize how concepts unfold in the video and how learners transition between them. We reuse four relation types—sequence, inclusion, association, and similarity—to organize COIs: sequence captures the order in which COIs are presented or visited; inclusion marks local COIs embedded in a larger conceptual episode; association links COIs that co-occur in the same context or arise from revisiting the same concept across segments; and similarity groups COIs that are close in content or gaze patterns and often contrasted or compared in instruction. This relational network embeds each COI in a broader conceptual and behavioral structure, enabling instructors to trace the teaching path via sequence relations or compare related concepts across segments via association and similarity.

\subsection{Integration of COI and Eye tracking}\label{sec:integration}

After defining COIs as our basic units of analysis, we now describe how to transform raw video content and eye tracking data into COI-based representations of learners’ behaviors for systematic analysis.

\subsubsection{Video-based Concept Extraction}

\textbf{Video preprocessing.} For multimodal MOOC video data, several precise steps are necessary, including speech recognition, slide extraction, and element recognition.

\noindent
\begin{itemize}
\item \textit{Speech recognition}. We separate the audio channel from the video and, for videos lacking subtitles, apply Automatic Speech Recognition (ASR)\cite{malik2021automatic} to obtain transcripts. For videos with subtitles, we directly extract the embedded text. All transcripts are stored with timestamps for concept extraction and alignment with slide segments.

\item \textit{Slide extraction}. 
When original slides are unavailable, we segment the video into frames and apply Shot Boundary Detection (SBD)\cite{kar2024video} to locate major content changes, which serve as natural transition points for automatic slide extraction.

\item \textit{Element recognition}. After extracting slides, element recognition captures visual information. Text elements (including handwritten notes) are parsed via Tesseract OCR\cite{sporici2020improving} to identify key concepts. A CNN model\cite{kattenborn2021review} trained on 240 annotated frames distinguishes figures, tables, equations, code, and handwritten notes, and stores their timestamps and positions in JSON format.
\end{itemize}

\textbf{Concept extraction.} Accurate concept extraction from multimodal videos is central to building a clear knowledge framework. We leverage GPT-4V \cite{achiam2023gpt} to efficiently extract concepts, using transcribed speech as the primary input and designing prompts that incorporate slide text and layout (e.g., heading hierarchy, bullet lists, formulas, figures) as priors to link visual structure to concepts. Guided by a Chain-of-Thought model \cite{lyu2023faithful}, we progressively fuse these text sources into a hierarchical, temporally annotated concept framework in JSON format, where each node is a concept identifier associated with one or more lecture time intervals. We then perform relationship extraction, identifying \textit{sequence}, \textit{association}, \textit{similarity}, and \textit{inclusion}, following Zhou's work \cite{zhou2024conceptthread}.

\textbf{Concept alignment.} We align abstract concepts with concrete visual representations to obtain a concept-structured MOOC video. We first derive textual descriptions by visually analyzing non-text elements, e.g., turning trends in figures into text. We then integrate visual features (position, size, color, highlighting) and temporal features (start time, duration) of all slide elements. This semantic, visual, and temporal information is organized as context prompts for GPT-4V, producing a refined alignment between visual elements and concepts along the video timeline. For each concept identifier, we thus obtain (i) an \emph{instructional episode}—the set of video and audio segments in which the concept is presented—and (ii) a set of concept-driven AOIs—the screen regions where the concept is visually instantiated. Notably, a single instructional episode may span non-adjacent segments, breaking traditional contiguous-area paradigms. This approach facilitates a flexible, multilayered representation of the flow of concepts, where each concept, tied to specific time intervals and visual cues, reflects the dynamic nature of instructional delivery.

\subsubsection{Eye-tracking-based COI Mapping}

Eye tracking then bridges these concept-based video representations and learners' cognition by attaching learner-side gaze traces to the above instructional episodes to instantiate COIs.

\textbf{Eye tracking preprocessing.} We recorded eye movements using a Tobii Pro X2-30 remote eye tracker at 30~Hz on 1920$\times$1080 displays. Each session started with a 9-point calibration; sessions with unsatisfactory calibration were recalibrated or discarded. Raw gaze data were processed using Tobii’s I-VT (Identification by Velocity Threshold) pipeline\cite{olsen2012tobii}. Binocular gaze samples were combined, smoothed with a short moving median filter, and brief gaps in the data were interpolated, while longer gaps were preserved for blink detection. Based on angular velocity, samples exceeding a velocity threshold were labeled as saccades, and contiguous non-saccade samples above a minimum duration were classified as fixations, with very short fixations discarded. This procedure yielded three basic eye-movement events for each learner---fixations, saccades, and blinks.

\textbf{Spatiotemporal eye tracking mapping.} The learner's gaze must be aligned with the COI definition in both spatial and temporal dimensions. We operationalize visual cues as concept-driven AOIs, i.e., screen regions that visually instantiate a concept. For each learner and each concept identifier, we construct a \emph{concept-related gaze trace} by selecting gaze events whose timestamps fall within the concept's instructional episode and whose screen positions intersect any of its concept-driven AOIs; events outside AOIs during the episode are marked as GAPs. To focus on meaningful gaze, transient movements such as very brief glances (e.g., $<0.05$~s) are filtered out, leaving stable fixation patterns. Combining the concept identifier, its instructional episode, and the corresponding concept-related gaze trace instantiates COIs as concept-indexed spatiotemporal segments of gaze behavior, linking the ``where'' and ``when'' of gaze to the ``what'' of cognitive focus\cite{just1980theory}.


\subsection{COI-based Learner State Analysis}\label{sec:learnerstate}

COI-based learner state analysis characterizes how learners engage with COIs over time. COI sequences capture temporal, spatial, and quantitative patterns of engagement, while representative features describe the quality of learning on each COI and support the inference of higher-level cognitive states.

\subsubsection{COI Sequence Generation}

Building on the defined COIs and their capture, we first derive COI sequences that summarize each learner's trajectory across multiple COIs in MOOC videos. A series of COIs forms a time-ordered sequence, providing detailed records of each COI's learning order, quantity, duration, and frequency. For a learner $i$, the learning process in a video containing multiple levels of concepts can be represented as:
\begin{equation}
\begin{aligned}
\mathrm{Learner}_i = \{&COI_{1-1-1}, LST_{1-1-1}, LET_{1-1-1},\\
                      &COI_{1-1-2}, LST_{1-1-2}, LET_{1-1-2}, \ldots\}
\end{aligned}
\end{equation}

Here, $\mathrm{LST}$ and $\mathrm{LET}$ denote the start and end times of learner $i$'s engagement with each COI. To ensure the reliability of the sequences, brief jumps are excluded, and only segments formed by continuous fixations are used. In this way, the COI sequence provides a spatiotemporal path model that describes how the learner transitions from one cognitive object to another throughout the video, reflecting the evolution of the learner's state with respect to the concepts.

\subsubsection{COI Learner-state feature Mining}

Concept sequences show what learners focus on and when, but do not reveal deeper cognitive states directly. Eye tracking captures rich gaze metrics that have been linked to underlying processes~\cite{holmqvist2011eye,dengRevieweyetracking2023}, but these low-level measures are difficult for instructors to interpret. 
\edit{Drawing on the learner-state features identified through 
our literature review (\ref{sec:litreview}) and user interviews 
(\ref{sec:interview}), we operationalize five key features at 
the COI level, aligning low-level gaze metrics with 
interpretable learning constructs across intra-concept 
engagement and inter-concept integration.}


\textbf{Intra-concept engagement.} We model \feature{Attention}, \feature{Cognitive Load}, and
\feature{Interest} as gaze-based latent states and operationalize each at the COI level using established eye tracking indicators.

\edit{\feature{Attention} is defined as allocating cognitive 
resources to the current task while suppressing task-irrelevant 
thoughts. While fixation-based metrics are a common AOI-level 
proxy~\cite{holmqvist2011eye}, we adopt peak saccade velocity, 
an established indicator of attentional engagement and 
vigilance that is suited to the COI aggregation 
scale~\cite{di2013saccadic,bachurina2022multiple,chuikova2024eye}. 
\toolName{} accordingly uses mean peak saccade velocity within 
each COI window as the primary indicator.}

\edit{\feature{Cognitive Load} denotes the processing burden 
when understanding a concept, especially peaks from high 
information density. Pupil diameter is widely recognized as a 
reliable proxy for cognitive load and processing 
effort~\cite{van2018pupil,mathot2018pupillometry,krejtz2018eye,duchowski2018index}; 
\toolName{} uses baseline-corrected mean pupil diameter within 
each COI to quantify processing intensity.}

\edit{\feature{Interest} is defined as a learner's subjective 
attraction to, and willingness to invest effort in, a COI 
beyond task-driven fixation. Blink suppression around salient 
events has been associated with interest and sustained 
engagement~\cite{nakano2009synchronization,smilek2010out,maffei2018spontaneous,hollander2022extracting,ranti2020blink}; 
\toolName{} uses blink suppression relative to an individual 
baseline as a core indicator and combines it with COI coverage 
and revisit behavior to construct a composite interest feature.}

\textbf{Inter-concept integration.}
To capture how learners engage with the broader course structure and pace, we further analyze \feature{Preference} and \feature{Synchronicity}.

\edit{\feature{Preference} is a learner's tendency to devote 
disproportionate attention to particular concepts or information 
types. Fixation duration and dwell-time distribution have been 
shown to reflect information preference and depth of 
processing~\cite{shimojo2003gaze,krajbich2010visual,hu2020moocsubtitles,mu2019learners}; 
\toolName{} computes, for each COI, a dwell-time surplus by 
comparing its share of dwell time with its share of 
presentation duration.}

\edit{\feature{Synchronicity} denotes the temporal alignment 
between learner behavior and instructional delivery. Gaze-based 
``with-me-ness'' has been shown to predict learning outcomes in 
video-based 
settings~\cite{sharma2014me,madsen2021synchronized,madsen2022cognitive,liu2023using}; 
\toolName{} encodes each moment on the concept-level timeline 
as same, ahead, behind, or outside, and aggregates these into a 
synchronicity distribution per COI.}

\subsubsection{Computation of Learner-state features}
Each learner-state feature is computed per learner at the COI level and normalized by that learner’s overall behavior. For physiological signals such as pupil size and blink rate, we estimate a per-learner baseline over the full video and express COI-level values as deviations from it. \feature{Preference} and \feature{Synchronicity} are defined as relative proportions or differences within the learner’s distribution over all COIs. These concept-level scores can then be aggregated across learners to estimate each concept’s relative impact and to compare how a single learner behaves across concepts. For a learner and a COI $c$, we summarize the main gaze signals into five scalar scores:

{%
\setlength{\abovedisplayskip}{0pt}%
\setlength{\belowdisplayskip}{0pt}%
\setlength{\abovedisplayshortskip}{0pt}%
\setlength{\belowdisplayshortskip}{0pt}%

\feature{Attention}.  
Let $v_e$ denote the peak velocity of the $e$-th saccade within COI $c$, and $N_s$ the number of such saccades; higher $\mathrm{Att}_c$ indicates faster saccades and more active orienting during COI $c$.
\begin{equation}
  \mathrm{Att}_c = \frac{1}{N_s} \sum_{e=1}^{N_s} v_e,
\end{equation}

\feature{Cognitive Load}.  
Let $p_t$ denote the pupil size at sample $t$ within COI $c$, $N_p$ the number of valid samples, and $\mu$ be the learner's baseline pupil size over the whole video. We define $\mathrm{Load}_c$ so that positive values indicate pupil dilation relative to baseline, and thus higher processing demands.
\begin{equation}
  \mathrm{Load}_c = \frac{1}{N_p} \sum_{t=1}^{N_p} (p_t - \mu),
\end{equation}

\feature{Interest}.  
Let $\mathrm{BR}^{\mathrm{all}}$ be the learner's overall blink rate and $\mathrm{BR}_c$ the blink rate during COI $c$; the interest score is defined so that larger $\mathrm{Int}_c$ reflects stronger blink suppression and thus higher engagement with COI $c$.
\begin{equation}
  \mathrm{Int}_c = \mathrm{BR}^{\mathrm{all}} - \mathrm{BR}_c,
\end{equation}

\feature{Synchronicity}.  
Within the active time window of COI $c$, let $T_{\mathrm{same}}$ be the total fixation time on the current COI region and $T_{\mathrm{all}}$ the total fixation time on all regions. The synchronicity score indicates the fraction of fixation time aligned with the currently instructed concept.
\begin{equation}
  \mathrm{Syn}_c = \frac{T_{\mathrm{same}}}{T_{\mathrm{all}}},
\end{equation}

\feature{Preference}. 
Let $\mathrm{DT}_c$ and $\mathrm{DT}^{\mathrm{all}}$ denote the dwell time on COI $c$ and on all COIs, and $\Delta T_c$ and $\Delta T^{\mathrm{all}}$ the presentation duration of COI $c$ and of the full video. The preference score (dwell-time surplus) is defined so that positive $\mathrm{Pref}_c$ indicates proportionally more time on COI $c$ than expected from its presentation duration.
\begin{equation}
  \mathrm{Pref}_c
  = \frac{\mathrm{DT}_c}{\mathrm{DT}^{\mathrm{all}}}
    - \frac{\Delta T_c}{\Delta T^{\mathrm{all}}},
\end{equation}

}%

\begin{figure*}[tb]
  \centering 
  \includegraphics[width=0.96\linewidth]{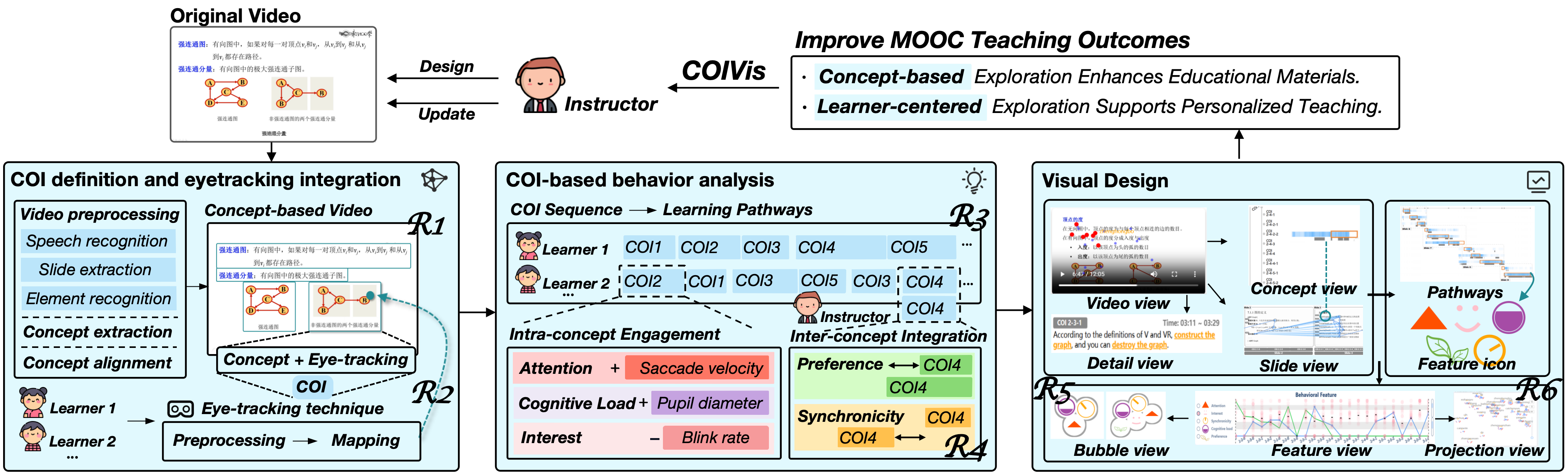}
  \caption{%
  	The pipeline of \toolName{} consists of three modules: COI definition and eye tracking integration, COI-based learner-state feature analysis, and visual analytics design.
  }
  \label{pipline}
  \vspace{-15pt}
\end{figure*}
\section{\toolName{}}\label{sec:coivis}

Section\edit{~\ref{sec:coi}} has introduced how to construct COIs from multimodal MOOC videos and eye tracking data, and how to quantify learning behaviors at the concept level. 
This section then presents our interactive visual analytics system, \toolName{}, which systematizes the above COI-based analysis pipeline and provides concept-level analytical support for instructors’ teaching diagnosis.

\subsection{System Overview and Workflow}\label{sec:overview}
Motivated by the design considerations DC1–DC3, we develop and implement an interactive visual analytics system, \toolName{} (Fig.~\ref{main}), to explore and analyze learners’ learning processes in MOOC videos. As illustrated in \edit{Fig.~\ref{pipline}}, the system is organized into three stages, each addressing one of the challenges CH1–CH3 introduced in Section\edit{~\ref{sec:intro}}.

In the preprocessing stage (corresponding to \textbf{CH1}), \toolName{} \delete{ingests}\edit{runs an offline pipeline, jointly verified by three authors, that transforms} multimodal course materials (MOOC videos, speech transcripts, and lecture slides) and eye tracking logs\delete{, and transforms them} into COI-ready data. 
It first constructs a concept map from the course content and aligns each concept to a concept time window and concept-driven AOIs on the video, thereby anchoring abstract concepts to concrete spatiotemporal positions (\textbf{R1}). 
The cleaned eye tracking data are then mapped onto these concept positions by assigning fixations that fall within the corresponding concept time window and concept-driven AOIs, yielding COI segments—spatiotemporal units of gaze behavior indexed by concepts (\textbf{R2}). 
\delete{This preprocessing pipeline is currently executed offline using the methods described in Section~\ref{sec:integration}, and the resulting concepts and mappings are jointly reviewed once by three authors to ensure the plausibility of key concepts and their spatiotemporal alignment.}


In the learner-state analysis stage (corresponding to \textbf{CH2}), \toolName{} constructs, for each learner, a COI sequence that encodes the order in which concepts are visited in the video, the dwell time on each concept, and any revisits (\textbf{R3}). Based on the definitions in Section\edit{~\ref{sec:learnerstate}}, the system then derives five learner-state features—Attention, Cognitive Load, Interest, Preference, and Synchronicity—from the eye tracking metrics associated with each COI, and normalizes these features with respect to the learner’s overall viewing behavior in the entire video, thereby highlighting relatively high or low values at the concept level. Taken together, the COI sequences and learner-state features provide two complementary perspectives on learners’ cognitive processes: one describes how learners engage within individual concepts, and the other reflects how they integrate and transition between concepts (\textbf{R4}).

In the visual analytics stage (corresponding to \textbf{CH3}), the curated COIs and learner-state features are loaded into the system via the \feature{Control panel}  (Fig.~\ref{main}(a)), and \toolName{} initializes a set of coordinated views, including the \viewname{Feature Panel}, \viewname{Main View}, \viewname{Projection View}, \viewname{Slide View}, and \viewname{Detail View}(Fig.~\ref{main}(b–f)). 
The \viewname{Feature Panel} uses a bubble-based design to summarize the five learner-state features for each concept and allows instructors to filter and rank COIs according to their diagnostic needs, helping them quickly locate concepts with abnormal patterns (\textbf{R5}). 
The \viewname{Main View} presents the hierarchical structure of concepts and visualizes learners’ aggregated learning states on these concepts; the \viewname{Detail View} provides content-level context and entry points to the original video segments. The \viewname{Slide View} displays the original lecture slides and embeds visual elements that encode the relationships between COIs. The \viewname{Projection View}, in turn, presents differences between learners from a class-level perspective and highlights subgroups or individuals with atypical learning patterns (\textbf{R6}).


\edit{Overall, \toolName{} turns teaching analysis into a concept-driven workflow, letting instructors first surface concepts with anomalous learner states and then drill down across the coordinated views to reach concept-level diagnoses.}

\subsection{Usage Scenario and Interaction Flow}\label{sec:usage}
To illustrate how the above components work together in practice, this subsection outlines a typical usage scenario based on our collaborations with instructors in the smart classroom.

\subsubsection{From Global Overview to Focal COIs}
After preprocessing a MOOC session and loading the corresponding COI dataset, the instructor selects the target course in the \viewname{Control Panel} (Fig.~\ref{main}(a)), and \toolName{} initializes the \viewname{Feature Panel}, \viewname{Main View}, \viewname{Projection View}, \viewname{Slide View}, and \viewname{Detail View} for that session (Fig.~\ref{main}). The instructor first follows the instructional sequence and browses slides step by step in the \viewname{Main View} (Fig.~\ref{main}(d)) and \viewname{Slide View} (Fig.~\ref{main}(f)), combining gaze heatmaps, behavioral curves, COI glyphs, and the global feature summary in the \viewname{Feature Panel} to gain an overall impression of how learning states evolve across the video. 
\delete{Simple range}\edit{Range} controls in the Feature Panel \delete{then support the transition from a ``global impression'' to ``identifying key issues'': the instructor filters}\edit{allow the instructor filter} COIs with, for example, consistently high \feature{Cognitive Load} or large variance in \feature{Attention}, upon which the \viewname{Main View} and \viewname{Slide View} automatically highlight the corresponding concepts, slides, and presentation regions, helping reveal where problematic concepts cluster in the course and with what types of content they co-occur (e.g., dense definitions or formula derivations).
At this point, the instructor typically forms preliminary hypotheses about why these COIs are difficult, which then guide subsequent analyses of learner groups and individuals.

\subsubsection{Drilling Down to Individual Learning Paths}
Once a set of focal concepts has been identified, the analysis shifts from ``which concepts are problematic'' to ``how these problems manifest in different learners,'' mainly via the \viewname{Projection View} (Fig.~\ref{main}(e)) and the single-learner mode. The \viewname{Projection View} embeds learners according to their COI sequences and five learner-state features and encodes these features in a compact glyph, 
\edit{enabling the instructor to select subgroups with specific patterns (e.g., low overall \feature{Synchronicity} or high \feature{Cognitive Load} on many COIs);}
selected learners are consistently highlighted or re-aggregated across the \viewname{Main View}, \viewname{Slide View}, and \viewname{Feature Panel}, where concept structures, behavioral curves, gaze distributions, and feature summaries are recomputed for that subgroup. For individual learners who require special attention, clicking the corresponding point in the \viewname{Projection View} activates single-learner mode, in which all views focus on that learner by showing their personal COI sequence, gaze traces on relevant slides, associated video segments in the \viewname{Detail View}, a path line of concept transitions in the \viewname{Control Panel}, and bubbles in the \viewname{Feature Panel} that indicate how their performance on each concept deviates from the class distribution.

\edit{Through this workflow, \toolName{} helps instructors identify both where the main problems lie and how they affect different learners.}

\begin{figure*}[t]
  \centering

  \begin{minipage}[b]{0.30\linewidth}
    \centering
    \includegraphics[width=\linewidth]{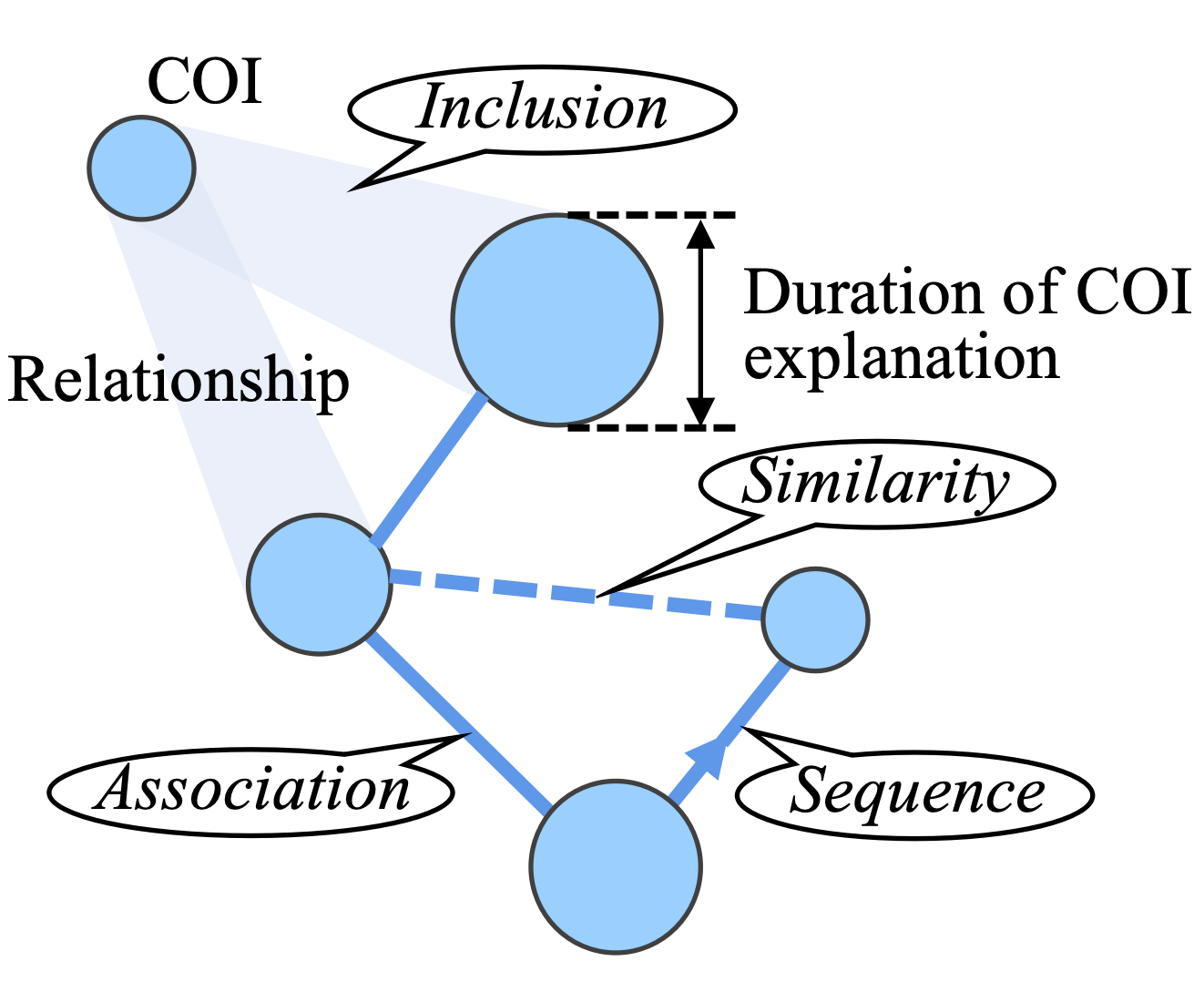}\\[-2pt]
    (a)
  \end{minipage}
  \hfill
  \begin{minipage}[b]{0.68\linewidth}
    \centering
    \includegraphics[width=\linewidth]{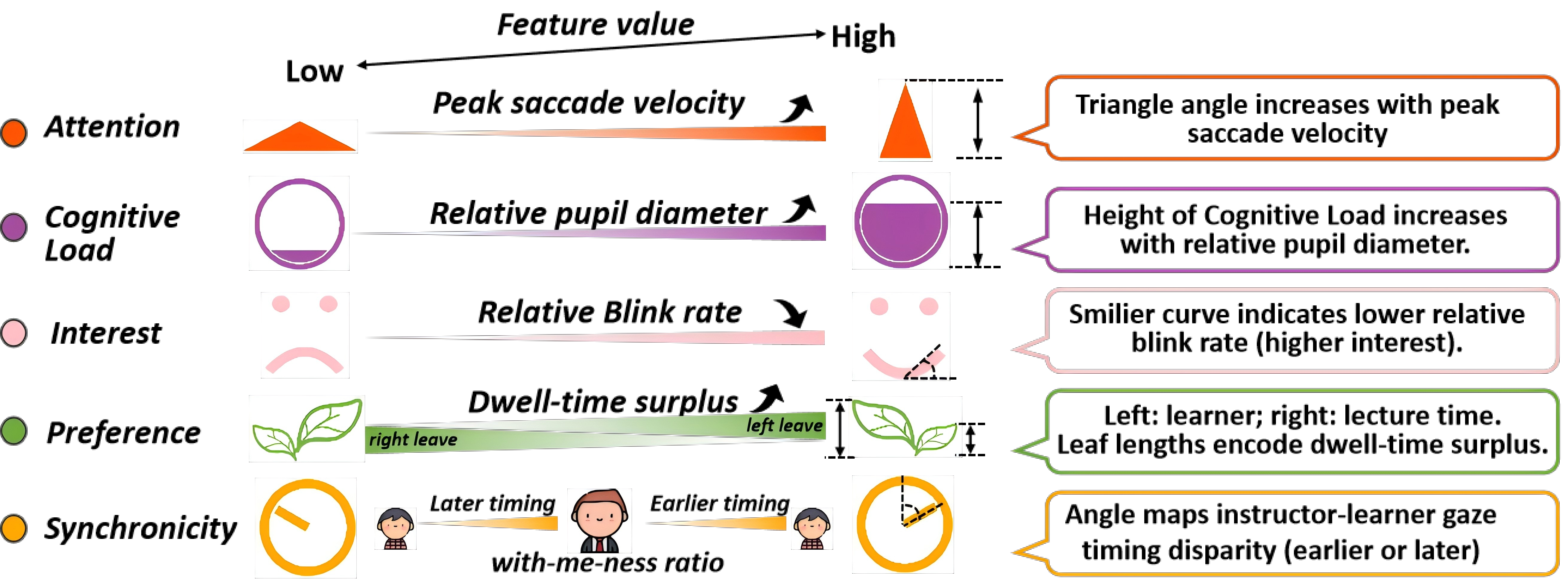}\\[-2pt]
    (b)
  \end{minipage}

  \caption{Visual design of COIs, relationships, and learner-state features.
  (a) shows how COIs are linked through four relationships; (b) illustrates how the five icons map to their respective learner-state features.}
  \label{COIrelations}
  \vspace{-5pt}
\end{figure*}

\subsection{Visual \& Interactive Design}\label{sec:visualdesign}
 Guided by principles of icon usability and redundant encoding~\cite{mcdougall2000exploring, isherwood2007icon, schroder2008effects, buhler2020universal, buhler2022designing} and classical visualization, interaction, and narrative design principles~\cite{shneiderman2003eyes, tufte1983visual, yi2007toward, segel2010narrative}, we design coordinated views, filters, and selections that summarize learner states at the concept level and enable transitions from course- and class-level overviews to focal concepts and, when needed, individual learner trajectories.

\subsubsection{Learner–Concept Interaction Visualization}

\toolName{} captures learner--COI interactions to represent MOOC learning processes. 
After importing course videos (Fig.~\ref{main}(a)), the \viewname{Main View} (Fig.~\ref{main}(d)) provides a course-level overview. Rather than traditional gaze plots or area-relationship diagrams, which hinder temporal comparison and pattern detection, we adopt a flow-based COI transition diagram (Fig.~\ref{main}(d1)) with multi-level nodes encoding conceptual hierarchies: vertical alignment reflects hierarchy, while horizontal positioning encodes temporal order. The expandable structure supports progressive detail exploration and granularity control. Following Shneiderman’s information-seeking mantra of “overview first, zoom and filter, then details-on-demand”~\cite{shneiderman2003eyes}, the \viewname{Main View} first presents global concept flows and then allows instructors to incrementally expand specific branches. 
COIs and their sequence, inclusion, association, and similarity relations are thus visually summarized as a concept-level transition diagram (Fig.~\ref{COIrelations}(a)).

Interviews (P3, P6) indicated that instructors prioritize cohort-level patterns to quickly detect common learning challenges. To support this, we aggregate behavior in two ways: (1) COI-specific bar heatmaps show gaze distributions, where color intensity denotes attention density and yellow temporal markers align instructional phases with attention trends; and (2) three color-coded stacked charts track real-time fluctuations in \feature{Attention}, \feature{Cognitive Load}, and \feature{Interest} across learning stages. Inspired by Tufte’s aligned, layered small multiples~\cite{tufte1983visual}, these vertically coordinated components enable at-a-glance comparison of temporal changes across concepts and features, revealing behavioral correlations (e.g., when \feature{Cognitive Load} rises, \feature{Attention} may decline or \feature{Interest} may shift) and informing timely instructional adjustments.

To concisely visualize five learner-state characteristics across COIs, we design semantically suggestive pictographic icons with joint color–shape encoding (Fig.~\ref{COIrelations}(b) shows their quantitative encoding and usage). Empirical studies on icon usability highlight semantic transparency and low visual complexity as key determinants of recognition efficiency~\cite{mcdougall2000exploring,isherwood2007icon,schroder2008effects}, so we draw on familiar metaphors from safety signage and everyday user interfaces and employ redundant encoding to enhance discriminability~\cite{buhler2020universal,buhler2022designing}. Specifically, \textbf{\textcolor[rgb]{1.00,0.00,0.00}{the red triangle}} represents \feature{Attention}, consistent with warning triangles in safety symbols~\cite{iso3864_2011,hellier2006handbook}; \textbf{\textcolor[rgb]{0.35,0.00,0.50}{the container icon}} for \feature{Cognitive Load} is partially filled with dark purple liquid, leveraging the metaphor of “load” filling limited capacity; \textbf{\textcolor[rgb]{1.00,0.41,0.71}{the pink smiley icon}} encodes \feature{Interest}, following smiley-face rating scales to capture subjective experience without textual labels~\cite{wong1988pain,hicks2001faces,stange2018effects}; \textbf{\textcolor[rgb]{0.00,0.80,0.00}{the green leaf}} denotes \feature{Preference}, drawing on everyday associations between leaves, growth, and “natural choice”; and \textbf{\textcolor[rgb]{0.80,0.80,0.00}{the yellow clock}} encodes \feature{Synchronicity}, using the ubiquitous metaphor of clocks as indicators of timing and coordination. These bubble icons in the upper \viewname{Main View}, support Yi et al.’s “encode” and “abstract/elaborate” interaction intents~\cite{yi2007toward}: they provide compact COI-level summaries that can be expanded into quantitative readouts when needed. While the three stacked charts show dynamic changes of \feature{Attention}, \feature{Cognitive Load}, and \feature{Interest}, the glyph-based design offers a holistic snapshot of learners’ overall performance on each concept; \feature{Synchronicity} and \feature{Preference} are summarized as single values rather than time series.

The COI transition view externalizes eye tracking data from the original stimuli, providing a structured visualization of learning processes. The associated \viewname{Slide View} (Fig.~\ref{main}(f)) improves semantic clarity by highlighting and aligning relevant content for selected COIs, supporting rapid content localization. The \viewname{Detail View} (Fig.~\ref{main}(b)) supplements dynamic and auditory cues by displaying original MOOC videos with instructors’ verbalized explanations, with eye movements temporally synchronized to video frames. Clicking a COI in the \viewname{Main View} jumps to the corresponding video segment, linking abstract concepts to concrete instructional context.

\begin{figure*}[t]
  \centering

  \begin{minipage}[b]{0.32\linewidth}
    \centering
    \includegraphics[width=\linewidth]{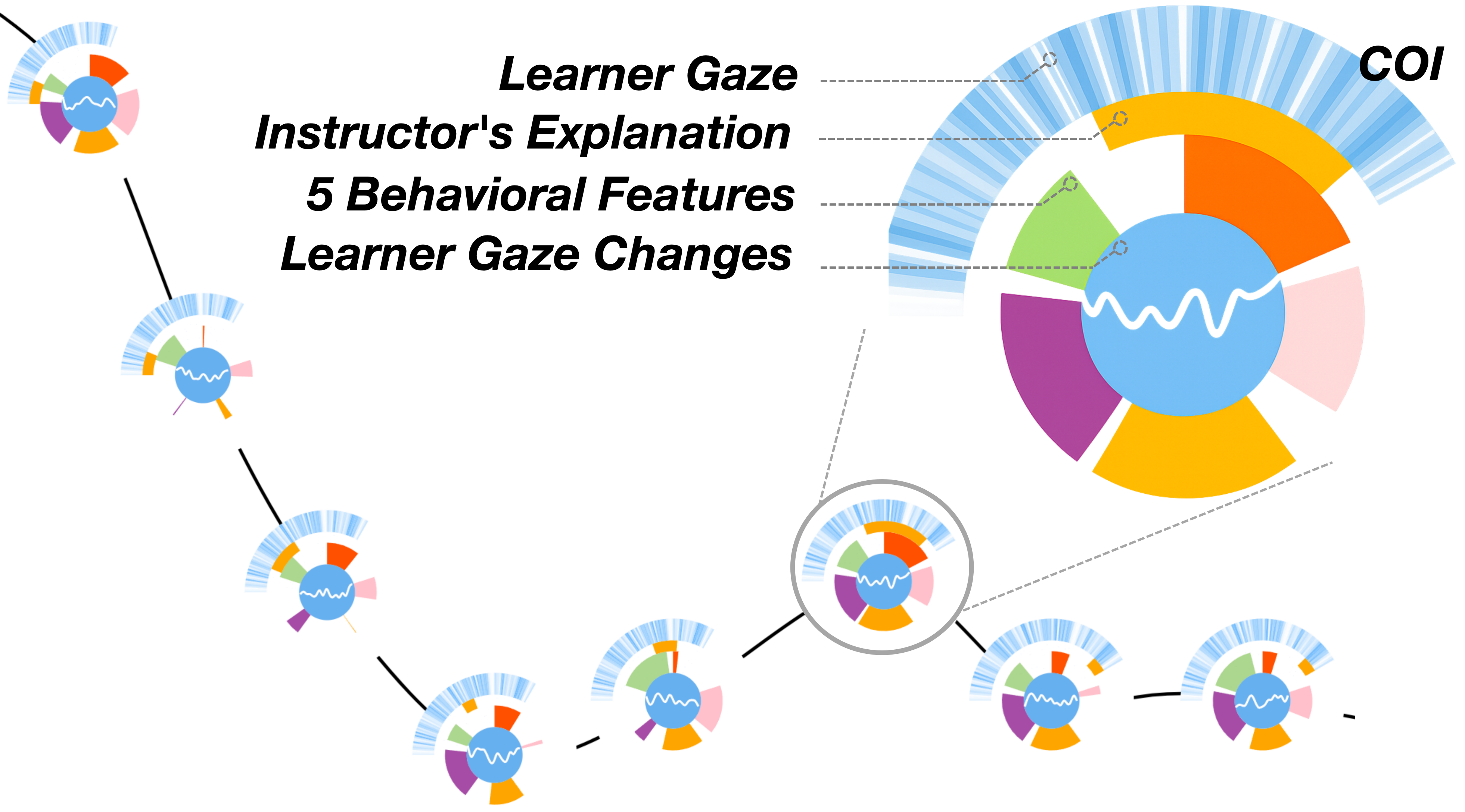}\\[-2pt]
    (a)
  \end{minipage}
  \hfill
  \begin{minipage}[b]{0.32\linewidth}
    \centering
    \includegraphics[width=\linewidth]{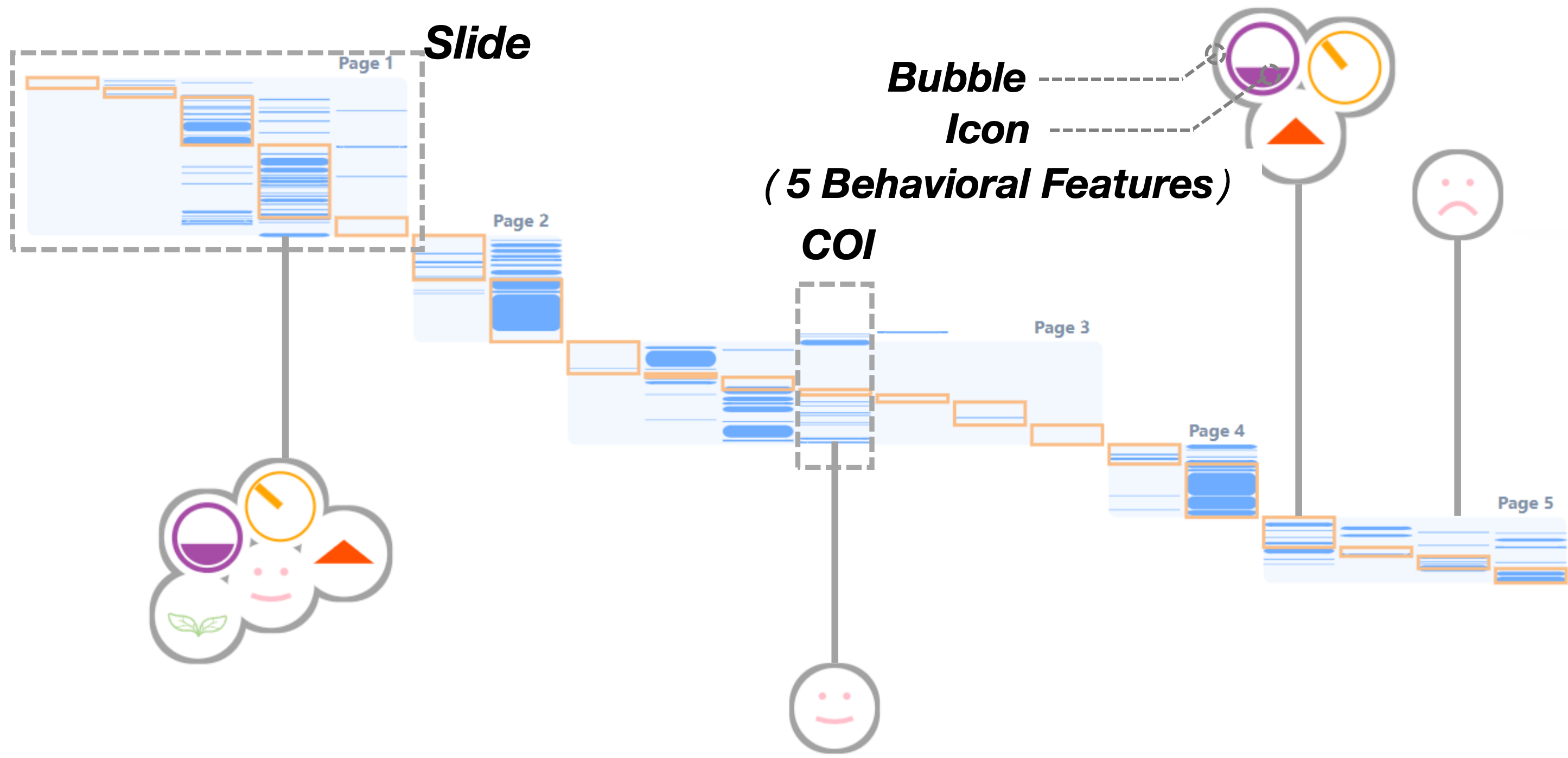}\\[-2pt]
    (b)
  \end{minipage}
  \hfill
  \begin{minipage}[b]{0.32\linewidth}
    \centering
    \includegraphics[width=\linewidth]{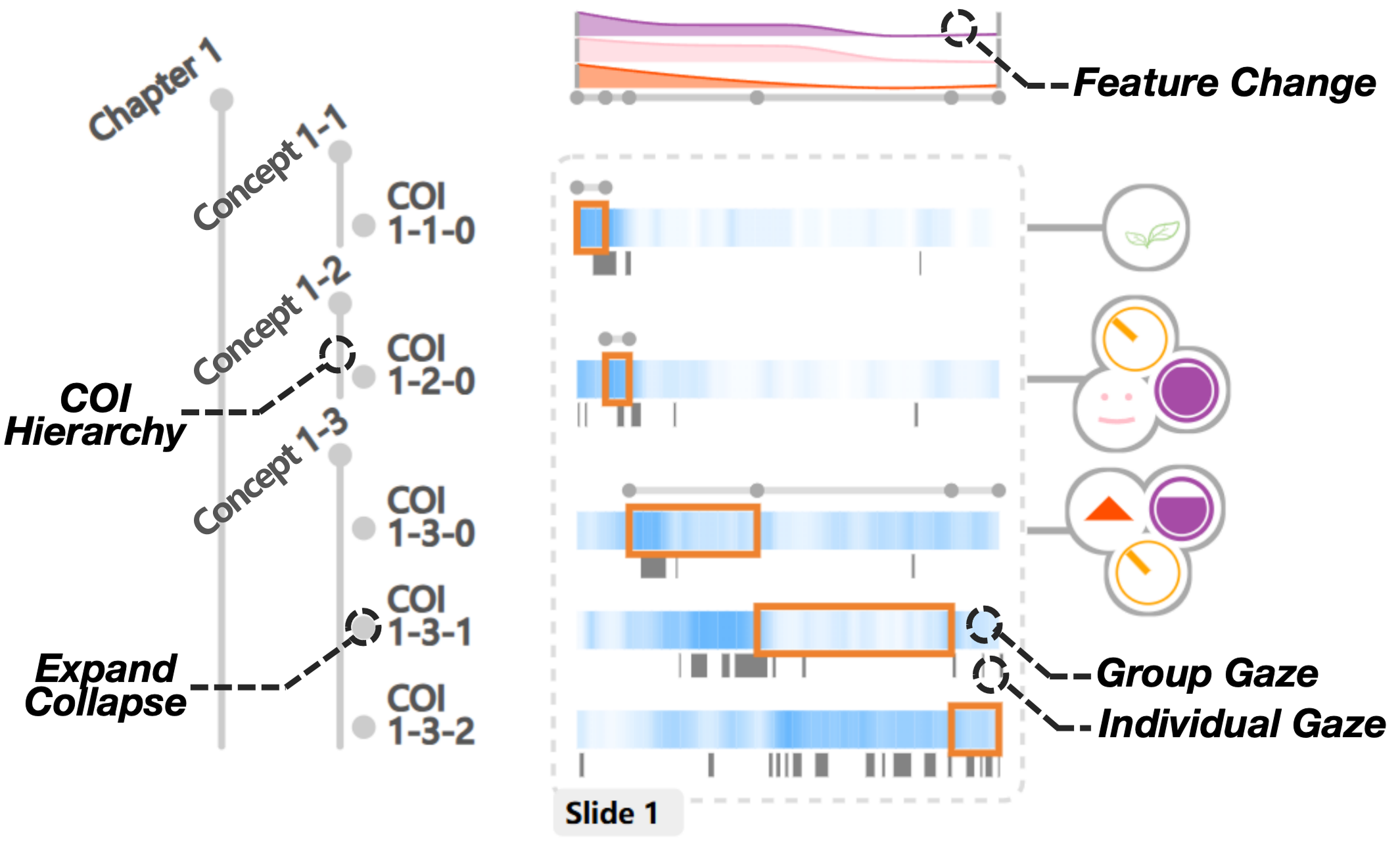}\\[-2pt]
    (c)
  \end{minipage}

  \caption{(a) and (b) represent alternative designs of COI pathways and learner-state features. (c) shows our final design.}
  \label{alternative}
  \vspace{-15pt}
\end{figure*}

\subsubsection{Instructor-driven Exploratory Visualization}

We further support instructor-driven exploration that moves from passive monitoring to evidence-based instructional decisions. 
\edit{The \viewname{Feature Panel} (Fig.~\ref{main}(b))} displays learner-state feature distributions across learners for each COI and operationalizes Yi et al.’s “filter” and “select” intents~\cite{yi2007toward}: two sliding windows on five learner-state dimensions help identify anomalous concepts (e.g., widespread high \feature{Cognitive Load} or large performance discrepancies) that warrant closer inspection. 
Once 
\edit{flagged through these range selections in the \viewname{Feature Panel}}, 
\edit{anomalous learner-state features} are highlighted in the \viewname{Main View} using an ``icon + bubble'' design
\edit{, with the corresponding features appearing as stacked icons in a bubble extending from each COI and providing} 
contextual explanations. 
Following Tufte’s guidance on preserving alignment and avoiding unnecessary complexity~\cite{tufte1983visual}, bubbles are placed in a consistent region next to each COI to maintain a clean, legible hierarchy.

During the design process, we evaluated two alternative options (Fig.~\ref{alternative}). The initial design (Fig.~\ref{alternative}(a)) used an eye-shaped glyph with a pupil ring encoding learner-state features and an eyebrow-shaped heatmap for attention, but heavy reliance on color and curved layouts hindered quick feature recognition and temporal accuracy. The second alternative (Fig.~\ref{alternative}(b)) adopted an “icon + bubble” design for rapid recognition and space efficiency, yet suffered from irregular bubble placement and unclear hierarchical relationships among COIs. Our final design (Fig.~\ref{alternative}(c)) retains the “icon + bubble” metaphor while implementing a scalable COI hierarchy with right-aligned bubbles, prioritizing visual consistency and hierarchical flexibility over strict horizontal temporal continuity. This reflects a Shneiderman-style trade-off~\cite{shneiderman2003eyes} between a stable, legible overview and focused details-on-demand.

To move from group trends to individual behaviors, the \viewname{Projection View} (Fig.\edit{~\ref{main}}(e)) embeds learners in a two-dimensional space based on their COI sequences and learner-state features. We first encode each learner’s COI sequence using Doc2Vec~\cite{article2014} and reduce the resulting vectors with t-SNE~\cite{van2008visualizing}; distances are then adjusted using Euclidean differences in learner-state features so that learners with similar profiles lie closer together. Instructors can select which features influence the embedding and tune their weights via sliders, instantiating Yi et al.’s “reconfigure” and “encode” interactions~\cite{yi2007toward}. Each learner is represented by a glyph inspired by Wu and Qu~\cite{wu2018multimodal}: three segments around the outer ring indicate values of \feature{Attention}, \feature{Cognitive Load}, and \feature{Interest}, while the central circle encodes \feature{Synchronicity} by partitioning the proportions of COIs viewed on time, ahead of, or behind the instructor, with darker colors indicating that the learner is ahead and lighter colors indicating that the learner lags behind. These glyphs are dynamically linked to other views, with panning and zooming used to mitigate overlap. Selecting a glyph activates single-learner mode across all views, displays a path line in the \viewname{Control Panel} tracing the learner’s concept transitions, and uses interactive bubbles to highlight performance deviations relative to the group. This drill-down from overview to focus and then to details follows Shneiderman’s overview–zoom–details pattern~\cite{shneiderman2003eyes}, while Yi et al.’s “connect” intent~\cite{yi2007toward} is realized by tightly linking the projection, main view, slide view, and detail view, enabling personalized guidance within a class-level context.

\section{Evaluation}\label{sec:eval}

 We conducted two case studies and a user-feedback interview based on real-world MOOC videos to evaluate its usefulness and effectiveness.

\subsection{Case Study}\label{sec:casestudy}

This section presents two case studies illustrating how instructors (P3 and P5) used \toolName{} to explore learners’ engagement while interacting with MOOC videos. The study was conducted in a smart classroom through controlled individual sessions. Each workstation (1920×1080) was equipped with a Tobii Pro X2-30 eye tracker, calibrated for each learner to ensure data accuracy. 
Learners viewed instructional videos in a quiet, distraction-free environment using headphones, followed by a 15-minute on-site test \edit{comprising multiple-choice and short-answer questions} that covered nearly all the concepts presented in the video.
To eliminate potential language-related confounds, two Chinese-language MOOC videos from the China University MOOC platform\footnote{\url{https://www.icourse163.org/}} were selected, ensuring that the focus remained on learning behavior rather than language proficiency.

\textbf{Case 1. Concept-based exploration enhances educational materials.} In this case, the learning material was the chapter \textit{"Definitions and Terminology of Graphs"} from the \textit{"Data Structures"} course\footnote{\url{https://www.icourse163.org/course/NEU-1003734012?from=searchPage&outVendor=zw_mooc_pcssjg_}}, with a video length of about 12 minutes. As a theory-heavy course containing many abstract concepts, it was well-suited for analyzing and comparing learners’ understanding across different concepts. A total of 47 undergraduate learners (29 male, 18 female) from Computer Science and Digital Media programs, spanning multiple academic years, participated in the study. After careful validation and cleaning of the eye tracking data, 36 complete datasets were retained and imported into \toolName{} for further analysis.

\begin{figure*}[t]
  \centering
  \includegraphics[width=\linewidth]{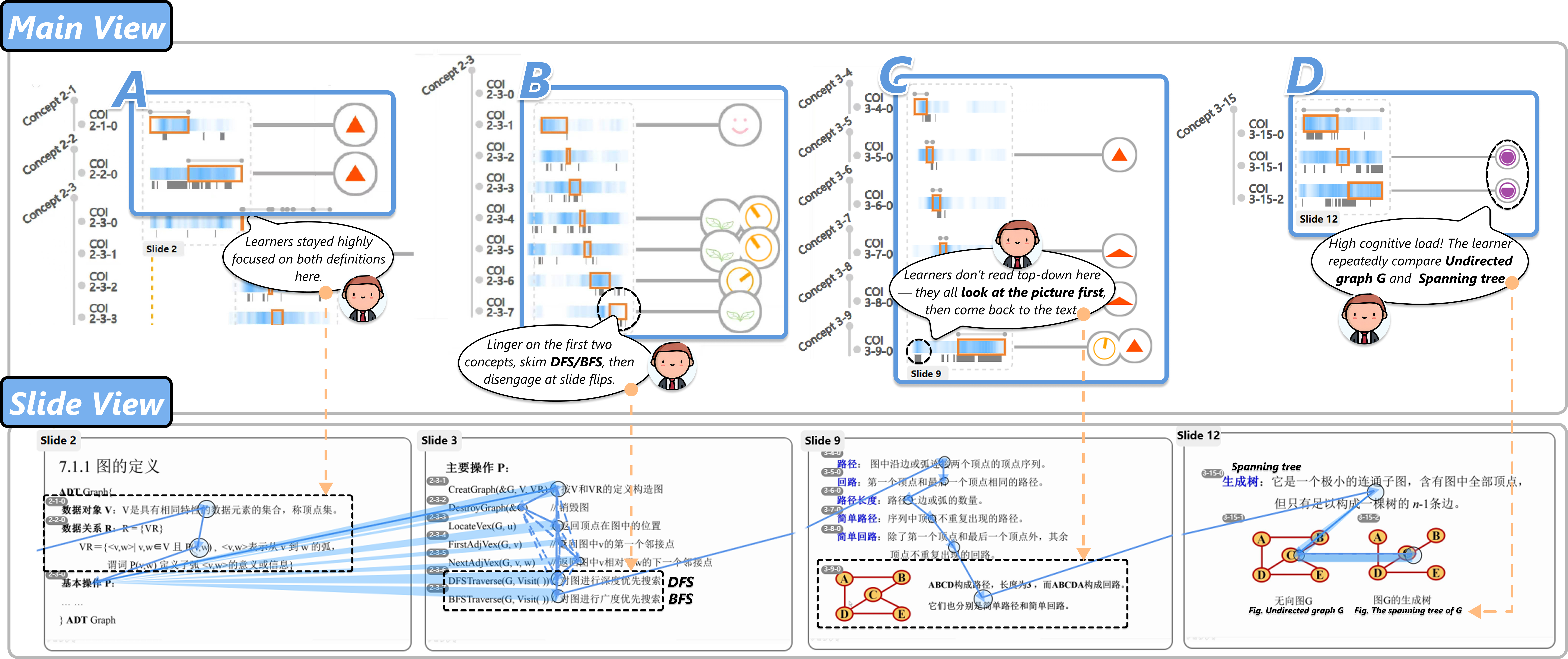}
    \caption{
      Four representative learning patterns observed by instructor P3 in the lecture \textit{Definitions and Terminology of Graphs}. Each panel pairs a Main-View summary (top) with the corresponding Slide-View snapshot (bottom) for the same concept window.
      (A) Sustained attention on the two textual definitions on Slide~2.
      (B) Uneven engagement across the algorithm list on Slide~3: learners linger on the first two concepts, skim DFS/BFS, and disengage after the slide transition.
      (C) On Slide~9, the visual example draws initial gaze before learners return to the surrounding text.
      (D) On Slide~12, high cognitive load and repeated back-and-forth comparison between the closely related concepts \textit{undirected graph}~$G$ and \textit{spanning tree}.
    }
  \label{fig:case1}
\vspace{-15pt}
\end{figure*}

\textit{Instructional Analysis.} 
P3, a senior Data Structures professor, used \toolName{} to analyze learners’ conceptual understanding and refine instructional video design to improve learning outcomes. P3 conducted a detailed analysis of each slide in sequence based on the teaching content. On \textbf{Slide 2}, the slide presents three concepts in text form. Heatmap data (Fig.~\ref{fig:case1}A) revealed that learners exhibited high levels of focused gaze on all three, indicating sustained attention. Upon reviewing the synchronized playback function in \toolName{}, P3 noted that the instructor used the mouse cursor to follow the text closely while explaining, dynamically highlighting key points. This pattern was consistent with P3’s impression that the revised explanation helped some learners focus more on the key steps of the concept, although he noted that stronger evidence would be needed to substantiate this effect.

P3 then analyzed \textbf{Slide 3} and observed that FirstAdjVex (COI 2-3-4) and NextAdjVex (COI 2-3-5) triggered similar ``leaf'' and ``clock'' bubbles (Fig.~\ref{fig:case1}B), indicating that learners spent significantly more time on these concepts than the instructor’s explanation, with their learning progress lagging behind. In contrast, for DFSTraverse (COI 2-3-6) and BFSTraverse (COI 2-3-7), learners spent much less time watching and progressed ahead of the instructor. P3 explained that while the first two functions are conceptually foundational, their theoretical presentation makes them harder to grasp quickly. The latter two functions, by contrast, are familiar traversal methods, so learners paid less attention and were still processing the previous two functions during the instructor’s explanation. Furthermore, P3 noticed a pattern of ``Page flipping disengagement'' on Slides 4, 5 and 6, where the learners' attention dropped after prolonged theoretical explanations.

A notable pattern emerged on \textbf{Slide 9}, where many learners ignored the expected top-to-bottom reading order, initially focusing on the only image (COI 3-9-0) before returning to the text (Fig.~\ref{fig:case1}C). Heatmap data confirmed that the image drew more attention than the surrounding theoretical concepts. When explaining concept COI 3-9-0, a clear re-engagement phenomenon occurred. P3, curious about this, reviewed the original video and found that the instructor said, ``Let's look at an example together'' while explaining COI 3-9-0. P3 speculated that using examples helped make abstract concepts more relatable, enhancing learner engagement. P3 observed that during the explanation of COI 3-9-0, learners' gaze shifted between related concepts like ``simple paths'' and ``simple circuits'' as the instructor referenced them.

On \textbf{Slide 12}, P3 observed that learners exhibited a high cognitive load (Fig.~\ref{fig:case1}D), aligning with cognitive load theory (CLT)\cite{plass2010cognitive}, which states that working memory is limited when processing new information, and cognitive load increases as learning progresses. He also noted frequent gaze shifts between ``undirected graph G'' (COI 3-15-1) and ``spanning tree of graph G'' (COI 3-15-2). Reviewing the video, P3 found that since this was a theoretical lecture, the instructor's brief explanation of generating a spanning tree likely led to learner confusion, prompting repeated comparisons between the two concepts. This back-and-forth viewing reflects a form of ``self-regulated learning''\cite{zimmerman2011self}, as learners sought to construct coherence between related yet insufficiently scaffolded ideas.

\textit{Instructional Implications.} The insights uncovered through \toolName{} provided P3 with concrete directions for improving instructional design. The key impacts include: (1) Employ synchronized visual-verbal techniques (e.g., cursor signaling) to maintain attention and enhance concept integration. (2) Integrate visual representations and concrete examples at key cognitive inflection points to support comprehension. (3) Segment content and adjust pacing to prevent disengagement after prolonged theoretical sections. (4) Strengthen instructional scaffolding between interrelated concepts helps resolve learner confusion and supports improved comprehension.

\begin{figure*}[tb]
  \centering 
  \includegraphics[width=\linewidth]{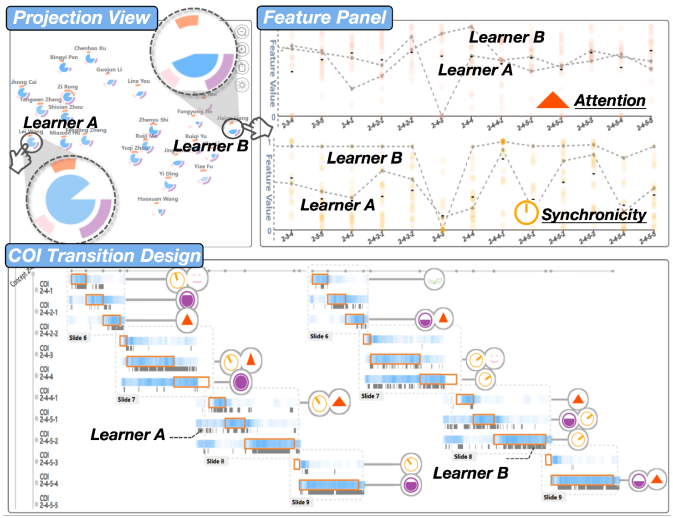}
  \caption{%
    Two selected learners are shown in the projection view. Their Attention and Synchronicity over a video segment are depicted as lines in the feature panel, with learning paths and performance visualized via COI transitions and bubble charts.
  }
  \label{case2}
\vspace{-15pt}
\end{figure*}

\textbf{Case 2. Learner-centered exploration supports personalized teaching.} We selected the \textit{"Regression Problem and Its Performance Evaluation"} chapter from the \textit{"Introduction to Deep Learning"} course\footnote{\url{https://www.icourse163.org/course/HIT-1206320802?from=searchPage&outVendor=zw_mooc_pcssjg_}} for Case 2. Unlike Case 1, which focused on pure theory, this lesson combines theory and application, explaining the computational process of various regression analysis evaluation methods. 
The study involved 36 undergraduates (21 male, 15 female) from Computer Science or Digital Media Technology programs. After data cleaning, valid data from 28 learners were retained.

\textit{Instructional Analysis.} P5, an experienced professor in machine learning and deep learning, was particularly interested in personalized instructional strategies. Using \toolName{}, P5 aimed to analyze individual learners' engagement patterns to support their diverse needs and learning styles. Initially, using \toolName{}'s projection view, P5 identified two learners with distinct engagement profiles. Learner A displayed a larger, light-blue sector, indicating high overall engagement but a tendency to lag behind the instructor’s pace. In contrast, Learner B’s smaller, darker-blue sector suggested a predisposition to grasp content ahead of the narration. \viewname{Feature panel} analysis confirmed this pattern, revealing that Learner A had lower attention and synchronicity scores across concepts, whereas Learner B maintained consistently higher synchronicity and relatively higher attention. Additional context identified Learner A as a first-year undergraduate new to the subject, while Learner B was a third-year learner preparing for graduate entrance exams. 
Quiz data confirmed these differences.
Upon reviewing the detailed learning pathways presented in Fig.\ref{case2}, P5 noted distinctive patterns.  Learner A followed the instructor’s sequential presentation of COIs but consistently lagged behind, indicating cognitive struggle or limited prior knowledge. In contrast, Learner B stayed ahead of the instruction, frequently revisiting earlier concepts non-linearly and focusing initial attention on visual representations. This gaze pattern led P3 to speculate that, in this lesson, the learner relied more on visual representations (e.g., diagrams) than on textual explanations when organizing information; however, P5 emphasized that this was only a tentative interpretation based on eye tracking data and would need further evidence from additional sources. P5 observed Learner A’s distraction during COI 2-4-4-1, which aligned with an incorrect quiz response. Despite normal gaze behavior in COI 2-4-3, Learner A still answered two related questions incorrectly, suggesting that regression loss analysis’s complexity may challenge novices even when visually engaged. In contrast, Learner B performed well, missing only one question despite moderate fixation. Notably, Learner B’s gaze was more dynamic, frequently shifting between explanation and calculation on Slide 7. P5 suggests that Learner B’s flexible, integrative learning approach could benefit instructors and peers.

\textit{Instructional Implications.} Based on the analysis, P5 and we identified key educational insights for personalized instruction and learning facilitation: (1)Adjust the teaching pace and style based on course types (e.g., foundational, review, or application) and prepare different versions of instructional videos (e.g., "detailed," "fast-paced," or "example-focused") to meet individual learner needs. (2)Integrate visual aids and graphical content to facilitate intuitive understanding and deeper cognitive processing, especially for visual learners. (3)Use analytics to promptly detect and address engagement lapses or confusion, providing timely intervention to maintain learner focus and clarity.

\subsection{User-feedback Interview}\label{sec:userfeedback}
In Section\edit{~\ref{sec:interview}}, we conducted comprehensive interviews with 8 instructors (P1-P8) to obtain their perspectives on the research requirements. To validate the implementation outcomes, we subsequently re-engaged the same cohort of instructors (P1-P8) for evaluations. 

\subsubsection{Interview design}
To collect systematic feedback from 8 instructors, our work implemented a three-phase interview protocol. Firstly, comprehensive orientation sessions were conducted to demonstrate \toolName{}' functional architecture and operational mechanisms. Subsequently, all participating instructors engaged in authentic classroom observations within smart learning environments. Finally, an interview in the form of a questionnaire were conducted, which systematically captured multi-dimensional evaluations regarding: (1) \textit{Learner-center exploration}, (2) \textit{Teaching feedback}, (3) \textit{Description of learning behaviors}, and (4) \textit{Visual and interactive experience}. 

\subsubsection{Interview results}
\begin{figure}[tb]
  \centering 
  \includegraphics[width=\linewidth]{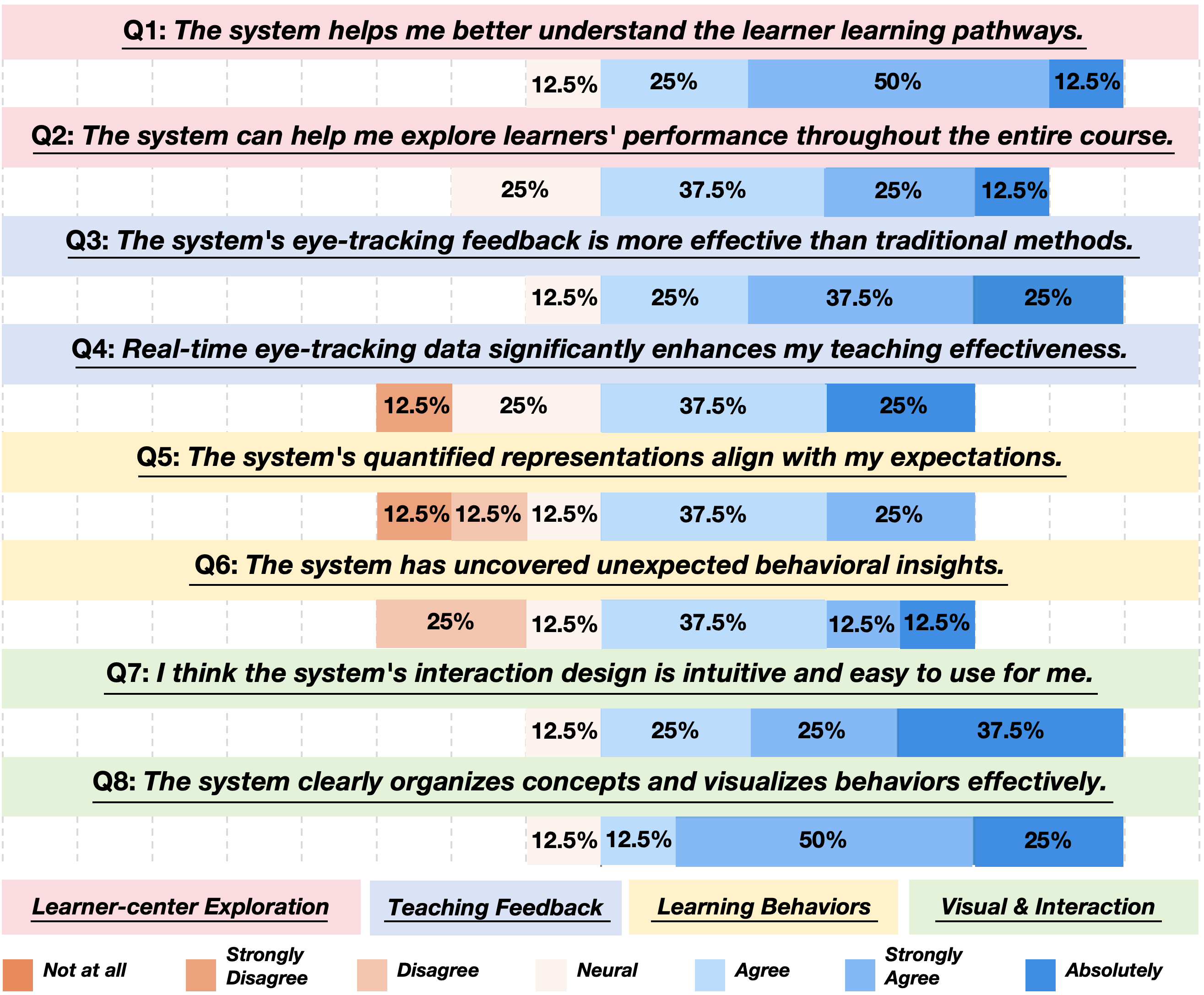}
  \caption{%
The results of Q1-Q8 in our questionnaire.
Agree scores to the right,
Neutral or Disagree to the left. No participant selected Not at all.  }
  \label{interview}
\vspace{-15pt}
\end{figure}

As shown in Fig.\ref{interview}, instructors provided favorable evaluations of \toolName{}, which received particularly positive recognition for \textit{Learner-center exploration} and \textit{Visual-interactive experiences}. While some instructors acknowledged the potential utility of eye tracking data in monitoring learners' attention levels during instruction, they simultaneously questioned its capacity to effectively assess knowledge acquisition. Furthermore, instructors raised critical inquiries regarding the identification of supplementary behavioral metrics that could more comprehensively characterize learning engagement states.

\textbf{Learner-center exploration.} P4 remarked, \textit{" \toolName{} exhibited exceptional technical proficiency in video pre-processing and concept extraction. The extracted concepts aligned closely with my prior expectations."} P1 emphasized its strengths in revealing learners’ conceptual transitions: \textit{"The system meticulously captures learners’ learning pathways across both spatial and temporal dimensions. This capability not only facilitates macro-level trend analysis of group dynamics but also enables granular examination of individual behavioral patterns."} However, P3 noted a key limitation: \textit{"The system’s capacity is confined to analyzing a limited set of concepts within a single MOOC video. In practice, educators are often more interested in assessing learner performance across an entire course or a curriculum sequence."} 

\textbf{Teaching feedback.}  
Instructors indicated that by integrating an accurate COI capture mechanism with eye tracking data, the system overcomes the inherent limitations of traditional methods, such as clickstream analyses or self-report questionnaires, in terms of objectivity and real-time data collection. This advancement provides educators with a more robust and accurate feedback mechanism for classroom instruction.  
Furthermore, P3 emphasized the system’s potential for practical application, stating: \textit{"If real-time eye-movement data from learners could be synchronized with pre-uploaded instructional videos during classroom sessions, it would significantly enhance the system’s applicability and pedagogical impact."}

\textbf{Description of learning behaviors.}  Instructors reported that certain behavioral indicators, such as \feature{Attention} shifts during explanations of complex concepts and heightened \feature{Interest} when new ideas are introduced reflect accumulated teaching expertise. Additionally, P7 noted unexpectedly that\textit{ "learners' learning pace should ideally synchronize with the instructor's delivery. However, data feedback reveals that after a new slide appears, learners typically first quickly scan picture information before focusing on text, gradually aligning their attention with the lecture content. During this period, their gaze tends to linger more on picture information."}

\textbf{Visual and interactive experience.} The participating instructors consistently appraised the visual design and interaction of \toolName{}. P7 particularly commended the bubble chart implementation, remarking, \textit{"The symbolic representation through spherical aggregation of multi-dimensional features significantly enhances visual pattern recognition efficiency through cognitive chunking mechanisms." }
Regarding interaction design,  P2 and P4 emphasized that \toolName{} provides multidimensional analytical tools enabling seamless transition between macro-level class performance trends and micro-level individual learner analytics, thereby substantially improving data exploration flexibility and operational convenience.
However, the evaluation also identified areas for improvement. P8 noted that \textit{"The current system's initial dashboard exhibits high information density that may induce cognitive overload for novice users during initial adoption phases."}
Additionally, P8 recommended allocating more screen space to the Slide view to 
improve concept visibility without interference.

\edit{\section{Discussion and Future Work}}\label{sec:discussion}
\textbf{Scalability and Generalizability.} 
The current implementation of \toolName{} has been tested on MOOC videos of approximately 10 minutes in length, containing around 40 concepts, with 35 participants per session. While we estimate that the system can handle up to 100 learners and 100 concepts, scalability challenges persist. In the main view, expand–collapse functionality and adjustable sliders help manage information overload, but in the projection view, glyphs may overlap as the number of learners increases, reducing readability. Future iterations could incorporate sampling~\cite{wang2024multi}, filtering~\cite{wang2022context}, and progressive disclosure, where learners are initially represented as projection points and glyph details are shown on demand. Beyond data scale, our empirical evaluation is based on a small number of concept-dense, visually supported STEM MOOC videos and undergraduate learners from a single university, which constrains generalizability across age groups, disciplines, and learning contexts. 
True educational impact will require longitudinal studies that track conceptual development over entire courses~\cite{evans2023concept,chen2024effects} and deployment of \toolName{} as a real-time tool (e.g., a plugin for educational platforms) to support dynamic, instructor-driven analysis during live sessions~\cite{ally2019competency}. 
\edit{Equally important is how the insights from \toolName{} eventually shape instructors' teaching adjustments and, in turn, learners' outcomes. Investigating this question, which extends into instructional-design research, is a key 
direction for our future work.}
Extending beyond MOOCs to other concept-rich narrative domains, such as educational documentaries~\cite{zhang2025embedding}, instructional videos, and professional training content, is another important direction.

\textbf{Concept Modeling and Pipeline Reliability.}
Learners may hold varying interpretations and cognitive needs regarding the same concept~\cite{borghi2022concepts}, which complicates the standardization of concept definitions. \toolName{} currently lacks flexible concept segmentation interactions that would allow instructors to adapt concept granularity based on learner needs. Additionally, subtitles and facial regions---frequent visual focal points in MOOC videos~\cite{narimani2024extracting}---are not yet modeled, leaving open questions about whether attention to these areas reflects meaningful concept engagement or distraction. In practice, COIs and gaze-to-concept mappings are generated by the semi-automatic preprocessing pipeline in Section\edit{~\ref{sec:integration},} but still require manual inspection to correct missed or misaligned concepts. The pipeline currently runs offline; integrating it into \toolName{}, assessing its reliability, and providing lightweight editing tools (e.g., splitting/merging concepts, correcting mappings) would help instructors detect and fix automatic errors directly within the interface.

\textbf{Interpreting Eye-Tracking Signals.}
The same eye-tracking indicator can stem from different psychological processes, making it insufficient to pinpoint specific cognitive mechanisms. Moreover, \toolName{} currently maps each learner-state feature to a single eye-tracking indicator, which may not fully capture the complexity of learner cognition. 
\edit{The five features are therefore heuristic, literature-informed proxies rather than formally validated measures; metric--construct validation in this context is left as future work.}
Future work should explore combining multiple eye-tracking metrics and integrating physiological and behavioral data~\cite{cuve2022validation} 
\edit{, grounded in established frameworks such as Cognitive Load Theory and the Cognitive Theory of Multimedia Learning,}
to more accurately quantify learning behaviors and to validate the mapping between gaze-based features and learner states.
\edit{Combining gaze-based proxies with external measures such as self-reports, quizzes, and learning-gain tests is another key future direction.}

\edit{\textbf{Usability and Practical Value.}}
User feedback indicates that the main view has the steepest learning curve, whereas the other views are readily understood. Initially, users may find the main view overwhelming because it presents a large amount of information simultaneously. Future refinements could focus on progressive information layering~\cite{wang2023progressive} and adaptive complexity reduction to enhance usability for first-time users, while preserving the rich, multi-level exploration capabilities needed for expert instructors.
\edit{Beyond usability within \toolName{} itself, comparing its value against alternative tools is another important dimension. A key direction for our future work is therefore a controlled comparison study examining how efficiently and 
accurately instructors locate concept-level learning difficulties using \toolName{} versus conventional clickstream- or quiz-based analytics dashboards. Such a study would further demonstrate the practical value of eye-tracking-based analysis.}


\section{Conclusion}\label{sec:conclusion}

\edit{We present \toolName{},}
a visual analytics system designed for instructors to analyze how learners engage with MOOC videos from a concept-based perspective. Developed from a smart classroom experiment, \toolName{} utilizes eye tracking to examine learner behavior. Specifically, we define and extract COIs and employ eye tracking data to measure \feature{Attention}, \feature{Cognitive Load}, \feature{Interest}, \feature{Preference}, and \feature{Synchronicity}, providing a structured representation for learning MOOC videos. \toolName{} employs a narrative visual design integrated with icon-based encodings and rich interactions, allowing instructors to gain both macro-level trends and fine-grained individual insights into learners’ concept comprehension. Case studies and user-feedback interviews validate its effectiveness in bridging the gap between MOOC learning and learner feedback, making it a valuable tool for data-driven teaching. In future work, we plan to expand \toolName{} to make it work for broader educational contexts, evolving into a comprehensive smart education analytics platform with enhanced scalability, behavioral interpretation, and multimodal learning integration.

\bibliographystyle{abbrv-doi-hyperref}

\bibliography{template}

\end{document}